%% file: main.tex
\documentclass[a4paper,11pt]{article}
\usepackage{jcappub} 
\usepackage{lineno}

\usepackage[table]{xcolor}

\title{\boldmath The impact of Hawking radiation from primordial black holes on recombination and the Hubble tension}

\author{Adam Batten and Jeremy Mould}
\affiliation{Swinburne University,
PO Box 218, Hawthorn 3122, Australia}
\affiliation{ARC Centre of Excellence for Dark Matter Particle Physics (CDM) Australia}
\emailAdd{abatten@swin.edu.au}

\abstract{
Primordial black holes (PBHs) evaporate through Hawking radiation, emitting high-energy photons and ionising their surrounding environment. As contributors to the density of dark matter, $\Omega_C$, if 10$^{-18}$ M$_\odot$ PBHs are present in the Universe at the time of recombination, they delay the cooling process and move the surface of last scattering of the cosmic microwave background by a substantial amount. We perform recombination simulations using the software \texttt{recfast}, measure the effect of adding non-monochromatic PBH heating on recombination and calculate the PBH fraction of dark matter required to resolve the Hubble tension. We find that
 nominally a cosmic PBH energy density of $\Omega_\mathrm{PBH} \approx 10^{-3}~\Omega_C$ would cause an 8.9\% increase in the value of $\mathrm{H_0}$, enough to entirely reduce the tension between the early- and late-time observations. However,
grey body factors, which are unknowable to the extent that they depend on PBH spin and charge, and range from 0.1\% to unity, hinder an accurate correction to H$_0$ for a 
modified ionization history.

We note that $\mathrm{H}_0$ is extremely sensitive to $\Omega_\mathrm{PBH}$. A slight increase in $\Omega_\mathrm{PBH}$ (even a $3\times$ increase) leads to a substantial change in $\mathrm{H_0}$. We find an absolute upper limit of $\Omega_\mathrm{PBH} < 10^{-2}~\Omega_C$, beyond which the density of PBHs would cause a complete reionisation of the Universe at these redshifts.
Until relevant non-gravitational properties of the dominant dark matter species are ruled out, we suggest that the hypothesis that the ionisation history of the universe matches the thermal history of the standard $\Lambda$CDM cosmology is too precarious to hang the expansion rate on, and that it is better to measure H$_0$ locally at z $\lesssim$ 1.}

\begin{document}
\maketitle
\flushbottom
\input{blackh0.tex}

%% file: blackh0.tex



\section{Introduction}
The Hubble tension is the significant (currently $>$ 5$\sigma$) discrepancy between local observations of our Universe's expansion rate, H$_0$ (the present-day Hubble parameter) and the inferred value from early Universe measurements ($ z \approx 1100$). The conflicting measurements between the early- and late-time observations of H$_0$ lead to a fundamental (though modest) disagreement about the age of the Universe.

Reviews by \cite{DiValentino2021} and by \cite{Vagnozzi2023} summarize around a thousand papers attempting to find solutions to this discrepancy, but the Hubble tension remains without a generally agreed solution.
\cite{Perivolaropoulos2022} proposed that there might be an evolution in the properties of the supernova standard candle. However, this theory conflicts with the Dark Energy Survey data on supernovae \citep{Vincenzi2024} in the range it covers, 0.1 $<$ z $<$ 1.1. Even new multi-messenger techniques that use gravitational waves as `standard sirens' have systematic uncertainties that could cause problems in resolving the tension  \citep{Bulla2022}.

Early-time observations of cosmic microwave background (CMB) radiation at a redshift of $z \approx 1100$ from \cite{Planck2020} indirectly measure the present-day Hubble parameter to be $\mathrm{H}_0~=~67.37\pm0.55~\mathrm{km~s^{-1}~Mpc^{-1}}$. This measurement is derived from inferring the value of H$_0$ from the best fitting cosmological model ($\Lambda$CDM + inflation) to the CMB.

On the other hand, local Universe observations, such as measurements of galaxy distances using the cosmic distance ladder and Type Ia supernovae, calibrated using Cepheid variable stars, consistently measure a `high value' of H$_0 \approx 73~\mathrm{km~s^{-1}~Mpc^{-1}}$ \citep{Riess2022, Riess2024, H0DN}. Other late-time methods, including the tip of the red giant branch stars \citep{Scolnic2023,Dixon2025}; the Tully-Fisher relation \citep{Kourkchi2020, Schombert2020}, time-delays from strong lensing of supernovae \citep{Pascale2024}; and Type II supernovae \citep{deJaeger2020}; all tend to agree with a higher value of H$_\mathrm{0}$ than inferred from \cite{Planck2020}. 

In this paper, we explore one mechanism in particular---primordial black holes---as an ionisation and heating source in the early Universe and their effect on recombination and the observed value of H$_0$.

Primordial black holes (PBHs) are black holes (BHs) that may have formed in the inflationary or early radiation-dominated Universe ($z>3400$) \citep{Zeldovich1967, Hawking1971}. Overdense fluctuations in the primordial matter density of approximately 10\% \citep{Harada2013} would undergo gravitational collapse and form a PBH. PBHs, like all BHs are non-baryonic, meaning that they could constitute some fraction of the dark matter in the Universe. PBHs could have a wide range of masses due to the diverse range of potential formation mechanisms \citep{Hawking1971, Carr1974, Carr2021}. \cite{Gabadadze2024} describe a mechanism for producing PBH with masses of the order
 10$^{-16}$ M$_\odot$.  
 
These PBH then evaporate through Hawking radiation \citep{p77}, injecting a large number of energetic photons ($E \geq 13.6$ eV) back into the cooling Universe, reheating and ionising nearby neutral gas. PBHs with lower masses will evaporate faster and at earlier times than those with higher masses
\citep{Mosbech2022}. PBH with masses less than $M_\mathrm{PBH} < 10^{-19}~\mathrm{M_\odot}$ will evaporate entirely before the end of recombination, injecting $M_\mathrm{PBH}c^2$ energy back into the early Universe, reheating the cooling gas
\citep{Sanchis}. An example is 10$^8$--10$^9$ g ($\sim10^{-26}$--$10^{-25}$ M$_\odot$) PBHs, whose abundance is severely limited by the data on Big Bang Nucleosynthesis \citep{Boccia2025}. The PBHs constrained in \cite{Boccia2025} have masses that are six orders of magnitude lower than the mass range we consider in this work.

If evaporating PBHs exist in the early Universe at the relevant mass range, they will inevitably cause recombination to be delayed. The extra reheating from the ionising photons counteracts some of the cooling from the Universe's expansion. A delay in the recombination time, and thereby the CMB surface of last scattering, would lead to an \textit{increased} value of H$_0$ (see \ref{app:H0_H}). \cite{Hamidreza2024} showed that modifications to the ionisation history during recombination can significantly reduce the tension between early- and late-time measurements of H$_\mathrm{0}$.

In this work, we have used the \texttt{recfast} recombination code \citep{Seager1999, Seager2000, Scott2009} to measure the effect that evaporating PBHs have on the ionisation history of the Universe. We perform recombination simulations, adding in the heating and ionisation created by the high-energy photons emitted through Hawking radiation. If a small portion of the mass density of dark matter ($10^{-3}~\Omega_C$) is in the form of PBH, the extra heating they contribute would delay recombination enough to alleviate the entire Hubble tension. However, if the proportion becomes too high ($10^{-2}~\Omega_C$), a full secondary reionisation occurs, preventing the Universe from becoming neutral until all the PBHs have evaporated.

In \ref{sec:Recfast}, we describe the \texttt{recfast} software and the parameters we use to perform the recombination simulations. In \ref{sec:photoionrate}, \ref{sec:directionisation}, and \ref{sec:secondaryionisation}, we describe the physical processes involved in the PBH heating and our determination of the photoionisation rate. In \ref{sec:implementrecfast} we describe how we implemented the heating  and ionization into \texttt{recfast}. In \ref{sec:Results}, we present the results of our simulations and how they impact $\mathrm{H}_0$. Finally, in \ref{sec:Conclusions}, we summarise our results, discuss them and present conclusions.

\section{Methods}

\subsection{Recfast recombination code} \label{sec:Recfast}
To perform recombination simulations, we use the software \texttt{recfast} (version 1.5.0) \citep{Seager1999,Seager2000,Scott2009} and add PBH radiation to the program's thermal evolution.

The software \texttt{recfast} was developed to accurately model the recombination history of the Universe, and its transition from ionised to neutral. It solves a modified 3-level atom for hydrogen and helium simultaneously, and has additional corrections including full treatment of background cosmology and radiation.  

To run recombination simulations, we provide a set of initial cosmological parameters, and starting from redshift $z=10^4$, \texttt{recfast} determines the temperature history down to redshift $z=0$, and outputs a series of ionisation fractions. The relevant output includes the degree of ionisation, $x_e$, and the ionisation fractions of hydrogen, $x_\mathrm{H}$, and helium, $x_\mathrm{He}$ for each redshift step.

In \ref{sec:implementrecfast} we describe in more detail the process of modifying \texttt{recfast} to include  ionization and heating from PBHs. However, the idea is to calculate boost factors, $\Delta x_\mathrm{H}$ and $\Delta x_\mathrm{He}$, to modify the ionisation fractions of hydrogen and helium at each redshift step to account for the effect of PBH heating.

In all of our simulations we use the following cosmological parameters, $\Omega_b = 0.04$, $\Omega_C = 0.2$, $\Omega_\Lambda = 0.76$, $T_\mathrm{CMB}=2.725~\mathrm{K}$ and $Y_p = 0.25$ (the cosmic baryon density, cosmic dark matter density, cosmic dark energy density, the CMB temperature and the primordial helium abundance respectively). The exact input values and cosmology are not critical to this work; instead, our aim is to provide an estimate of the change in the redshift of recombination, $\Delta z$, hence, the change in $\mathrm{H}_0$.

\subsection{Photoionisation rate}\label{sec:photoionrate}

The process of heating and photoionising the gas from Hawking radiation, has two main channels, \emph{direct ionisation} and \emph{secondary ionisation}. Direct ionisation involves electrons directly absorbing photons emitted through Hawking radiation. Secondary ionisation is a process where electrons gain energy indirectly through Compton scattering and pair-production interactions with the high-energy photons ($E \geq 1~\mathrm{MeV}$).

\subsubsection{Direct ionisation}\label{sec:directionisation}
In direct ionisation the hard ($E \geq 1~\mathrm{MeV}$) Hawking radiation photoionises neutral atoms, and electrons carry kinetic energy away. This process directly heats the gas.

The number of neutral atoms that are ionised by photons in the time $\Delta t$ is given by $\Gamma_{\gamma,n} \Delta t$ where $\Gamma_{\gamma,n}$ is the photoionisation rate,
\begin{equation}
    \Gamma_{\gamma,n} = \frac{c}{\lambda_\mathrm{mfp}} n_\mathrm{photons} = \sigma_\mathrm{PE} ~n_0 ~n_\mathrm{photons} c \,.
    \label{eq:ionisation_parameter}
\end{equation}
Here $\lambda_\mathrm{mfp} = (\sigma_\mathrm{PE} n_0)^{-1}$ is the mean free path of photons, $\sigma_\mathrm{PE}$ is the photoelectric effect cross section, $n_0$ is the neutral atom number density, $n_\mathrm{photons}$ is the photon number density

The direction ionisation rate of a gas is fundamentally limited by the photoelectric effect cross section, $\sigma_\mathrm{PE}$, which is given by the Pratt-Scofield equation \citep{Pratt1960, Scofield1973,Hubbell1980}, 
\begin{equation}
    \sigma_\mathrm{PE} = Z^5 \left( \sum_{n=1}^{4} \frac{a_n + b_n Z}{1 + c_n Z} \left(\frac{E_\gamma}{E_e} \right)^{-p_n} \right) \sigma_\mathrm{T}\,,
    \label{eq;Pratt-Schofield}
\end{equation}
where $Z$ is the atomic number of the absorbing atom ($Z=1$ for H, and $Z=2$ for He), $E_\gamma$ is the energy of the photon in eV, $E_e$ is the rest mass energy of the electron ($E_e = m_ec^2 \approx $ 0.511 MeV).  
 The parameters $a_n$, $b_n$, $c_n$ and $p_n$ are presented in \ref{tab:Pratt_Schofield_params}. The Thompson cross section for an electron, $\sigma_\mathrm{T}$ is given by,
\begin{equation}
\sigma_\mathrm{T} = \frac{8\pi}{3} \left( \frac{e^{\,2}}{4\pi\,\epsilon_0 \,m_e\, c^{\,2}}\right)^2 \approx 6.6524\times 10^{-29}\,{\rm m}^2\,,
\end{equation}
where $\epsilon_0$ is vacuum permittivity, $e$ and $m_e$ are the charge and rest mass of the electron, respectively.

\begin{table}
    \centering
    \begin{tabular}{ccccc}
        \hline
         $n$ & $a_n$ & $b_n$ & $c_n$ & $p_n$  \\
         \hline
         1 & $1.6268\times10^{-9}$ & $-2.683\times10^{-12}$ & $4.173\times 10^{-2}$ & 1 \\
         2 & $1.5274\times10^{-9}$ & $-5.110\times10^{-13}$ & $1.027\times10^{-2}$ & 2 \\
         3 & $1.1330\times10^{-9}$ & $-2.177\times10^{-12}$ & $2.013\times10^{-2}$ & 3.5 \\
         4 & $-9.12\times10^{-11}$ & 0 & 0 & 4\\
         \hline
         & 
    \end{tabular}
    \caption{Parameters of the Pratt-Schofield equation for the photoelectric effect cross section shown in \ref{eq;Pratt-Schofield} \citep{Pratt1960,Scofield1973} as presented in \cite{Hubbell1980}.}
    \label{tab:Pratt_Schofield_params}
\end{table}

However, as shown in \ref{eq;Pratt-Schofield}, the cross section for the photoionisation is a strongly declining function with increasing energy of the incoming photon, $E_\gamma$
The photons emitted through Hawking radiation of PBH are very hard ($E\sim 100~\mathrm{MeV} - \mathrm{GeV}$) due to their high Hawking temperature, $T_\mathrm{Hawk}$ \citep{Page1976}, 
\begin{equation}
    T_\mathrm{Hawk} = \frac{\hbar c^3}{8\pi  k_\mathrm{B} G M_\mathrm{PBH}}\,.
\end{equation}
Here $M_\mathrm{PBH}$ is the mass of the PBH, and $G$, $k_\mathrm{B}$, and $\hbar$ are the universal gravitational, Boltzmann's, and reduced Planck's constant, respectively.

As a result, direct ionisation of the gas is a subdominant process. The primary process for heating the gas is through \emph{secondary ionisation} as described in \ref{sec:secondaryionisation}. We chose not to include the process of direct ionisation, and all heating arises from secondary ionisation.

\subsubsection{Secondary ionisation}\label{sec:secondaryionisation}
According to \cite{Slatyer2009}, high-energy photons ($>$ 1 MeV) that undergo Compton scattering and pair-production interactions have rapid energy losses. The timescale for these energy losses ($t_\mathrm{losses}\sim\lambda_\mathrm{mfp} / c$) is significantly more rapid than the expansion time ($t_\mathrm{expansion}\sim\mathrm{H}(z)^{-1}$) of the gas surrounding the PBH. Due to energy losses from Compton scattering and pair-production occurring on timescales significantly faster than the expansion of the gas, these energy losses must be deposited into the surrounding gas, thereby thermally heating the gas. 

This means that all the photon energy from an evaporating PBH is eventually transferred into the surrounding gas, even though the probability of direct ionisation is low.
Hence, we can assume that all the energy released from the evaporating PBH goes into heating and ionising the recombining gas (i.e $\Delta E_\mathrm{gas} = - \Delta E_\mathrm{PBH}$). We refer to this process of heating and ionising the gas though Compton and pair-production processes as \emph{secondary ionisation}.

If we assume that the ratio of baryons-to-dark matter is preserved in the 
vicinity of a PBH, then a PBH of mass $M_\mathrm{PBH}$ will have an environment of gas surrounding it of mass $M_\mathrm{gas}$,
\begin{equation}
    M_\mathrm{gas} = M_\mathrm{PBH}\frac{\Omega_b}{f_\mathrm{C,PBH} \Omega_C}\,,
    \label{eq:Mgas}
\end{equation}
where $\Omega_b$ and $\Omega_C$ are the cosmic baryon and cosmic cold dark matter densities, respectively and $f_\mathrm{C,PBH}$ is the fraction of cold dark matter in the form of PBHs.

By assuming the heat capacity of a monoatomic ideal gas ($3/2~k_\mathrm{B}T$ per mole), the change in internal energy of the gas, $\Delta E_\mathrm{gas}$, surrounding the PBH is as follows, 
\begin{equation}
    \Delta E_\mathrm{gas} = \frac{3}{2} N_\mathrm{moles} k_\mathrm{B} \Delta T_\mathrm{gas} =\frac{3}{2}  \frac{M_\mathrm{gas}}{\mu m_p} k_\mathrm{B} \Delta T_\mathrm{gas}\,,
    \label{eq:Delta_Egas_1}
\end{equation}
where $\Delta T_\mathrm{gas}$ is the change in temperature of the surrounding gas, $N_\mathrm{moles}$ is the number of moles in the gas, $m_p$ is the mass of a proton, and $\mu=0.7$ is the mean molecular weight in atomic mass units. 

As illustrated by \cite{Mould2025}, PBHs with a mass in the range
$M_\mathrm{PBH}$ = 10$^{-21}$--10$^{-19}\mathrm{M}_\odot$  will evaporate entirely before recombination ($z\approx1100$). If we consider these PBHs that will evaporate entirely before recombination, then  the energy release 
~over the time before recombination is the entire mass of the PBH. 
 
As a result, $\Delta E_\mathrm{PBH} = M_\mathrm{PBH} c^2$. Setting this equal to \ref{eq:Delta_Egas_1} gives us the following,
\begin{equation}
M_\mathrm{PBH} c^2 = \frac{3}{2}  \frac{1}{\mu m_p} M_\mathrm{PBH}\left(\frac{\Omega_b}{f_\mathrm{C,PBH}\Omega_C} \right) k_\mathrm{B}\Delta T_\mathrm{gas}\,.     
\end{equation}
After rearranging for $\Delta T_\mathrm{gas}$, we find that the change in temperature of the gas surrounding the PBH is \emph{independent of the mass} of the PBH, as shown here in \ref{eq:DeltaT},
\begin{equation}
\Delta T_\mathrm{gas} = \frac{2\mu m_p c^2}{3k_\mathrm{B}} f_\mathrm{C,PBH} \left(\frac{\Omega_C}{\Omega_b} \right) \approx 2.6 \times 10^{13} f_\mathrm{C,PBH}~\mathrm{K}\,,
\label{eq:DeltaT}
\end{equation}
assuming $\Omega_b h^2 = 0.0224$ and $\Omega_Ch^2 = 0.12$ \citep{Planck2020}. We do not need to assume a value of $h^2$ as it cancels in the equation. This $\Delta T_\mathrm{gas}$ is reduced by the change in $(1+z)$ over the lifetime of the PBH due to adiabatic expansion of the gas, and redshifting of the photons.  The change in temperature of the gas environment is independent of the PBH mass because a higher-mass PBH is accompanied by proportionally more gas, as shown in \ref{eq:Mgas}. This means that the increased energy emitted by a higher-mass PBH is spread out over a larger number of particles, which causes the same overall temperature change.
\subsubsection{Neutrino losses}
At the Hawking temperatures relevant to the mass range considered
(T$_H ~\sim$ MeV–GeV for M $\sim$ 10$^{-18}$ M$_\odot$), a black hole emits photons, electrons
and positrons, neutrinos, and -- above QCD threshold -- quarks and gluons.
Neutrinos interact negligibly with the recombination-era plasma
and carry away a large fraction of the total Hawking luminosity
without depositing any energy.
For a Schwarzschild black hole at these temperatures, the neutrino fraction
of the total emission is approximately 40-50\%
\citep{page76, jane90} 
Our 100\% deposition assumption
therefore overestimates the effective heating rate by roughly a factor of two;
correspondingly, the f$_{C,PBH}$ required to produce a given shift in z* is underestimated by a comparable factor. 

\subsubsection{Redshift-dependent deposition efficiency}
Even the electromagnetic component of the Hawking emission is not deposited
instantaneously. High-energy photons at z $\sim$ 1000 have mean free paths
that are not necessarily short compared to the
Hubble radius, and the fraction of injected electromagnetic energy that
actually goes into ionization
(as opposed to heating or excitation) is a strong function of both
redshift and photon energy. The
standard framework for handling this, developed by \cite{spf09}
and substantially refined in subsequent work
\citep{sl2016, dh2020}, 
provides explicit, tabulated deposition efficiency functions 
for all relevant channels. 

\subsubsection{Change in ionisation fraction due to PBH heating, $\Delta x$}
Now, instead of considering the change in temperature of the gas over the entire lifetime of a single PBH, we calculate the change in ionisation fraction ($\Delta x$) resulting from the additional heating caused by a PBH evaporation in a single timestep $\mathrm{d}t$. During the timestep $\mathrm{d}t$, we apply the conservation of energy equation ($\Delta E_\mathrm{gas} = - \Delta E_\mathrm{PBH}$). We apply this rule for the reasons described in \ref{sec:secondaryionisation}. The ionisation fraction may exceed one, but only for the duration of a Compton cooling time, which is very much shorter than a \texttt{recfast} timestep.

The change in energy of a PBH in time $\Delta t$ is given by
\begin{equation}
    \Delta E_\mathrm{PBH} = c^2 \frac{\mathrm{d}M_\mathrm{PBH}}{\mathrm{d}t}\Delta t\,,
\end{equation}
where the $\mathrm{d}M_\mathrm{PBH}/\mathrm{d}t$ rate is given by the Hawking mass loss equation \citep{Mosbech2022},
\begin{equation}
    \frac{\mathrm{d}M_\mathrm{PBH}}{\mathrm{d}t} = - \frac{\hbar c^4}{G^2M_\mathrm{PBH}^2} \alpha_0\,.
    \label{eq:HawkingRate}
\end{equation}
Here $\alpha_0$ is given by,
\begin{equation}
    \alpha_0 = \begin{cases}
c_1 + c_2M_0^{p} & M_0 < 5\times 10^{-16}~\mathrm{M}_\odot\\
2.011 \times 10^{-4} & M_0 \geq 5\times 10^{-16}~\mathrm{M}_\odot
\end{cases}
\end{equation}
where $c_1= -0.3015$, $c_2 = 0.3113$ and $p=-0.0008$, and $M_0$ is the initial mass of the PBH.

Hence, the change in energy of the gas surrounding the evaporating PBH in the timestep $\Delta t$ is
\begin{equation}
    \Delta E_\mathrm{gas} = \chi \Delta n_\mathrm{ion} = - c^2 \frac{\mathrm{d}M_\mathrm{PBH}}{\mathrm{d}t} \Delta t \,,
\end{equation}
where $\Delta n_\mathrm{ion}$ is number atoms that become ionised in $\mathrm{d}t$ and $\chi$ is the ionisation potential, $\chi$ is $13.595~\mathrm{eV}$ for H and $24.587~\mathrm{eV}$ for He. Note that because $\mathrm{d}M_\mathrm{PBH}/\mathrm{d}t$ is negative, the change in energy of the gas is positive. We define the quantity $\Delta x$ (ionisation boost factor) as the number of ionisations per atom,
\begin{equation}
    \Delta x = \frac{\Delta n_\mathrm{ion}}{n_\mathrm{gas}}\,,
\end{equation}
where the number density of the gas surrounding the PBH, $n_\mathrm{gas}$, is obtained by dividing \ref{eq:Mgas} by the mean particle mass $\mu m_p$,
\begin{equation}
    n_\mathrm{gas} = \frac{M_\mathrm{PBH}}{\mu m_p}\frac{\Omega_b}{f_\mathrm{C,PBH}\Omega_C}\,.
\end{equation}

Hence, we have the number of extra ionisations per atom in time $\Delta t$ due to an evaporating PBH,
\begin{equation}
    \Delta x = -\frac{1}{\chi} \frac{\mu m_p c^2}{M_\mathrm{PBH}} \frac{\Omega_C} {\Omega_b}\frac{\mathrm{d}M_\mathrm{PBH}}{\mathrm{d}t} f_\mathrm{C,PBH}  \Delta t\,.
    \label{eq:Delta_x}
\end{equation}
Notably, this equation primarily depends on the mass of the PBH, the cosmology of the simulation and the length of the timestep.

\subsection{Recfast Simulations}\label{sec:implementrecfast}

\definecolor{white}{HTML}{FFFFFF}
\definecolor{NoPBHColor}{HTML}{000000}
\definecolor{PBH1Color}{HTML}{0C5DA5}
\definecolor{PBH2Color}{HTML}{00B945}
\definecolor{PBH3Color}{HTML}{FF9500}
\definecolor{PBH4Color}{HTML}{FF2C00}
\definecolor{PBH5Color}{HTML}{845B97}

\definecolor{PBH6Color}{HTML}{74a9cf}
\definecolor{PBH7Color}{HTML}{74c476}
\definecolor{PBH8Color}{HTML}{fc8d59}

\definecolor{PBH9Color}{HTML}{cccccc}
\definecolor{PBH10Color}{HTML}{969696}
\definecolor{PBH11Color}{HTML}{525252}


\begin{table}

    \centering
    \begin{tabular}{cccccc}

         \hline
        Name & $f_\mathrm{C,PBH}$ & IMF & $M_\mathrm{max}$ & Relative $\Omega_\mathrm{PBH}$ & Colour  \\
        & &  & [$\mathrm{M_\odot}$] & & \\
         \hline
         NoPBH & 0.00 & - & - & - & \textcolor{NoPBHColor}{$\blacksquare$} \\
         PBH1 & $10^{-5}$   & $M^{-1}$ & $10^{-18}$    & $1\times$& \textcolor{PBH1Color}{$\blacksquare$} \\
         PBH2 &  $10^{-4}$  & $M^{-1}$ & $10^{-18}$    & $10\times$ & \textcolor{PBH2Color}{$\blacksquare$} \\
         PBH3 & $10^{-3}$   & $M^{-1}$ & $10^{-18}$    & $100\times$ & \textcolor{PBH3Color}{$\blacksquare$} \\
         PBH4 & $10^{-2.5}$ & $M^{-1}$ & $10^{-18}$    & $316\times$ & \textcolor{PBH4Color}{$\blacksquare$} \\
         PBH5 & $10^{-2}$   & $M^{-1}$ & $10^{-18}$    & $1000\times$ &\textcolor{PBH5Color}{$\blacksquare$} \\
         PBH6 & $10^{-5}$   & $M^{-2}$ & $10^{-18}$    & $1\times$ & \textcolor{PBH6Color}{$\blacksquare$} \\
         PBH7 & $10^{-4}$   & $M^{-2}$ & $10^{-18}$    & $10\times$ & \textcolor{PBH7Color}{$\blacksquare$} \\
         PBH8 & $10^{-3}$   & $M^{-2}$ & $10^{-18}$    & $100\times$ & \textcolor{PBH8Color}{$\blacksquare$} \\
         PBH9 & $10^{-6}$   & $M^{-2}$ & $10^{-17.5}$  & $0.1\times$ & \textcolor{PBH9Color}{$\blacksquare$} \\
         PBH10 & $10^{-5}$  & $M^{-2}$ & $10^{-17.5}$  & $1\times$ & \textcolor{PBH10Color}{$\blacksquare$} \\
         PBH11 & $10^{-4}$  & $M^{-2}$ & $10^{-17.5}$  & $10\times$ & \textcolor{PBH11Color}{$\blacksquare$}\\
         \hline
    \end{tabular}
    \caption{The parameters used in our \texttt{recfast} simulations. From left to right, the column names are: simulation name, the fraction of dark matter in the form of PBH ($f_\mathrm{C,PBH}$), the shape of the PBH initial mass function (IMF), the maximum mass of the IMF (the minimum initial mass is $10^{-20}~\mathrm{M}_\odot$ for all simulations), the PBH density relative to PBH1, and the colour on the plots.}
    \label{tab:sims}
\end{table}






We implement the heating effects of evaporating PBHs into \texttt{recfast} by modifying both the standard input of the program to include additional ionisation tables ($\Delta x_\mathrm{H}$) and the Fortran source code to recalculate the ionisation fraction at each timestep. 
We have also considered a more recent recombination code by \cite{dh2020} 
which goes deeper into the microphysics of energy deposition and ionization. Figure 1 compares the degree of ionization with 
\texttt{recfast} and \texttt{DarkHistory}.
For present purposes \texttt{recfast}
suffices.
\begin{figure}
\centering
\includegraphics[width=1.4\linewidth]{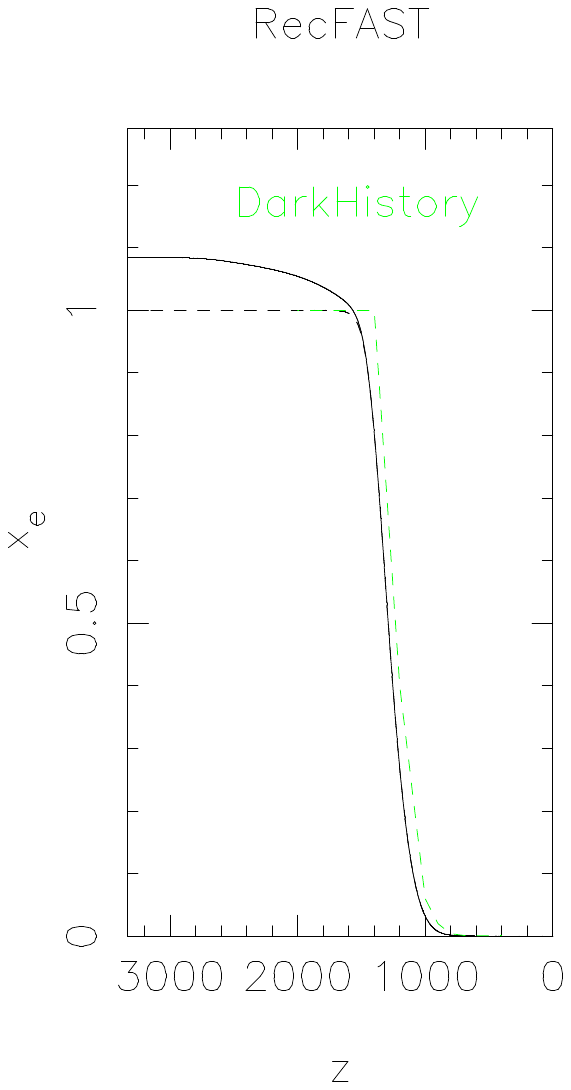}
\caption{The ionisation of hydrogen (dashed lines) as redshift decreases.  \texttt{DarkHistory} is shown in green.
	The solid line is the degree of ionization, x$_e$.}
\label{fig:darkhistory}
\end{figure}
\subsubsection{Calculation of $\Delta x_\mathrm{H}$ tables}\label{sec:calcdeltaHtables}
To obtain these $\Delta x_\mathrm{H}$ ionisation tables, we pre-compute the ionisation contribution from evaporating PBHs across the full redshift range of interest. For each timestep in our simulation, we integrate \ref{eq:Delta_x} over a mass distribution of PBHs, assuming $\chi = \chi_\mathrm{H} = 13.595~\mathrm{eV}$ for hydrogen. The mass distribution is discretized into 100 logarithmically spaced bins. We use 1000 linearly separated timesteps in redshift space between $z=10^4$ and $z=0$, giving a redshift step size of $\delta z_\mathrm{timestep} = 10$.

We use PBH mass range of $10^{-20}~\mathrm{M}_\odot$ up to a maximum cut-off, $M_\mathrm{max}$,  ($10^{-18}$ or $10^{-17.5}~\mathrm{M}_\odot$ see \ref{tab:sims}), because these are the PBH masses that will be actively evaporating around the time of recombination ($z \sim 1100$). Lighter PBHs would have already evaporated by this epoch, while more massive PBHs would still be largely intact and contributing negligibly to the energy injection. The choice of this mass window, therefore, maximises the impact on recombination while avoiding computational overhead from irrelevant mass ranges. We use an initial mass function (IMF) of the form,
\begin{equation}
    \frac{\mathrm{d}n_\mathrm{PBH}}{\mathrm{d}M} \propto M^{\alpha}\,,
\end{equation}
where $\mathrm{d}n_\mathrm{PBH}$ is the number of PBHs in the mass bin $\mathrm{d}M$.

It is important to state that because we use a distribution of masses, this means we are assuming PBHs are non-monochromatic in nature. Many existing constraints on PBH abundances such as \cite{Carr2010}, \cite{Poulter2019}, \cite{Carr2021} and \cite{Kohri2024} assume a monochromatic mass function (i.e. all PBHs have the same mass). A distribution of mass following an initial mass function (IMF) allows for a greater fraction of PBHs under the same constraint. There has been considerable research on extended PBH distributions such as \cite{Mould2025}, \cite{Musco2024} and \cite{Farooq2025}.   

This choice to end the mass range at $M_\mathrm{max}$ causes the heating due to PBH to abruptly truncate after recombination. In other words, once the most massive PBHs evaporate in our simulations, the amount of heating instantaneously drops to zero. A more comprehensive treatment would account for the temperature gradients and cooling processes involved due to this extra heating. We discuss this effect in \ref{sec:simparams} and \ref{sec:effect on recombination}.

Our modified \texttt{recfast} program reads the pre-computed ionisation boost factor, $\Delta x_\mathrm{H}$, and applies the PBH corrections and outputs the degree of ionisation, $x_e$, at each timestep,
\begin{equation}
    x_e = x_\mathrm{H,PBH}  + x_\mathrm{He,PBH}f_\mathrm{He}\,,
\end{equation}
where $f_\mathrm{He} = Y_p/(3.9715 (1-Y_p))$ is the number ratio of helium atoms, with $Y_p=0.25$. The corrected hydrogen ($x_\mathrm{H,PBH}$) and helium ($x_\mathrm{He,PBH}$) ionisation fractions after PBH heating are given as follows,
\begin{align}
    &x_\mathrm{H,PBH} = \min~(1, x_\mathrm{H} + \Delta x_\mathrm{H} )\,, \\
    &x_\mathrm{He,PBH} = \min~\left(1, x_\mathrm{He} + \frac{\chi_\mathrm{H}}{\chi_\mathrm{He}}\Delta x_\mathrm{H} \right)\,.
\end{align}
where $x_\mathrm{H}$ and $x_\mathrm{He}$ are the standard recombination ionisation fractions of hydrogen and helium calculated by the original \texttt{recfast} program, $\chi_\mathrm{H}=13.595~\mathrm{eV}$ and $\chi_\mathrm{He} = 24.587~\mathrm{eV}$ are the binding energies of hydrogen and helium respectively. We use a $\min$ function to ensure that the ionisation fractions remain physical and do not exceed unity.

\subsubsection{Simulation Parameters}\label{sec:simparams}
\begin{figure}
\centering
\includegraphics[width=\linewidth]{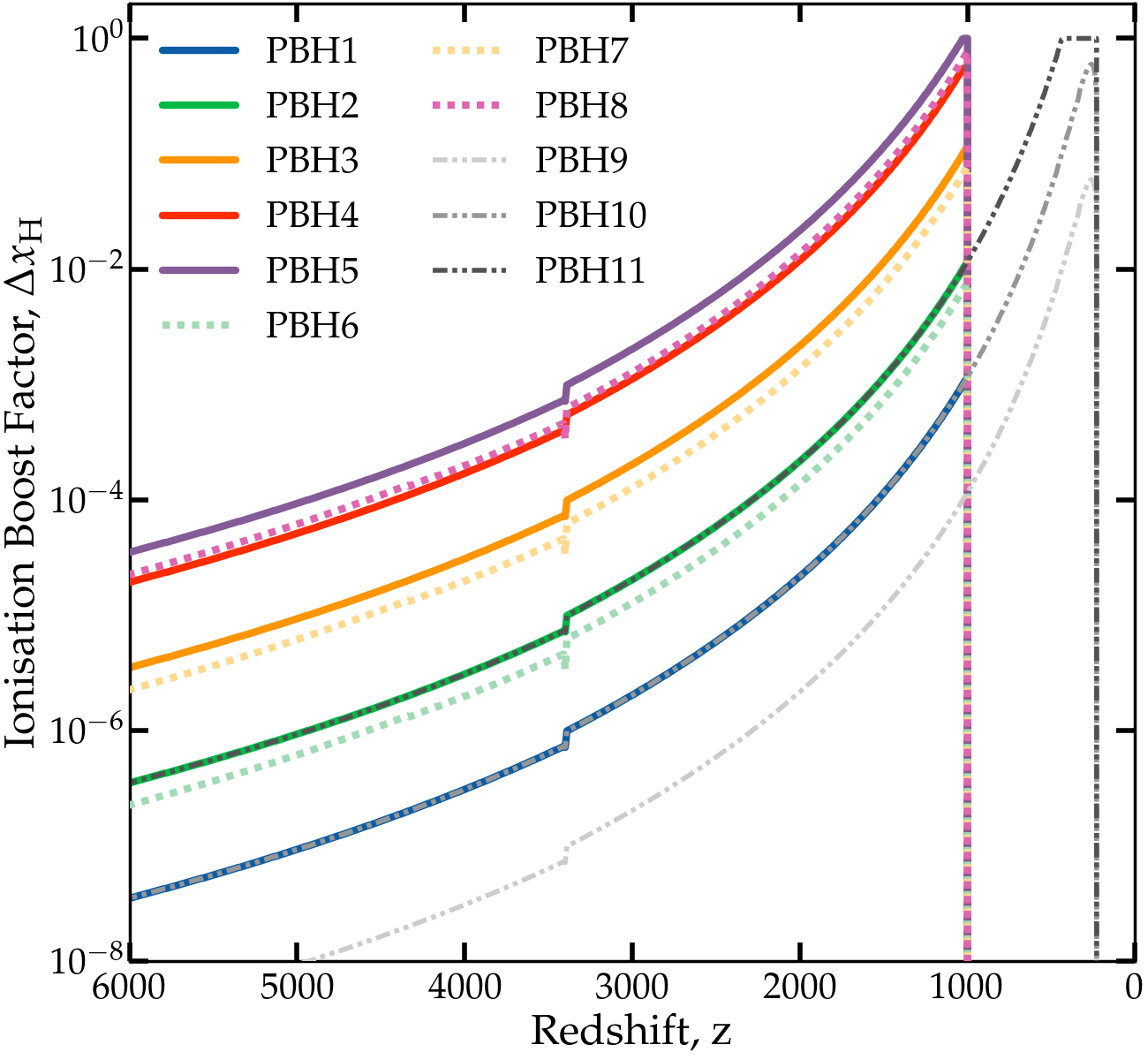}
\caption{The ionisation boost factor, $\Delta x_\mathrm{H}$, as a function of redshift ($z$) for the eleven PBH simulations. The solid lines are PBH simulations with an $M^{-1}$ IMF. Similarly, the dotted lines are the simulations with $M^{-2}$ IMF. The rapid drop-off in ionisation boost factor occurs at the timestep when all of the PBH in the simulation has evaporated.}
\label{fig:boost_factor}
\end{figure}

We performed twelve recombination simulations using our modified version of \texttt{recfast}: eleven simulations including PBH effects and one standard recombination simulation without PBHs for comparison (NoPBH). The parameters chosen for each simulation are detailed in Table 2.

Our parameter space explores variations in the fraction of dark matter in the form of PBHs ($f_\mathrm{C,PBH}$), the $M_\mathrm{max}$ and the slope of the IMF ($\alpha$). We have varied $f_\mathrm{C,PBH}$ across several orders of magnitude, from $10^{-6}$ to $10^{-2}$, to investigate how the abundance of PBHs affect the recombination history. We note that this value of $f_\mathrm{C,PBH}$ is the \emph{initial} fraction of PBH relative to dark matter (at redshift $z=10,000$)\footnote{We emphasise that throughout this paper, all references to $f_\mathrm{C,PBH}$ are referring to this initial abundance at $z=10,000$.}. As PBHs evaporate, this fraction decreases with time until it eventually drops to zero. The redshift at which all PBHs have evaporated is set by $M_\mathrm{max}$ in the IMF. We implement two distinct shapes of IMF: $M^{-1}$ (PBH1-5) and $M^{-2}$ (PBH6-11), representing different scenarios for formation and mass distribution of PBHs. An $M^{-1}$ IMF, as used in \cite{Mould2025}, puts an equal mass of PBH in each decade of mass bin. With an $M^{-2}$ IMF, we bias the IMF towards a higher number of lighter PBH. All PBH simulations use a lower bound on the IMF mass distribution of $10^{-20}~\mathrm{M}_\odot$ ($\sim 2\times10^{13}~\mathrm{g}$). For three of the $M^{-2}$ IMF simulations (PBH9-11), we have increased $M_\mathrm{max}$ to $10^{-17.5}~\mathrm{M}_\odot$ ($\sim 6\times10^{15}~\mathrm{g}$, up from $10^{-18}~\mathrm{M}_\odot$, $\sim 2\times10^{15}~\mathrm{g}$, for PBH1-8). This increase in $M_\mathrm{max}$, is to explore the consequences of delaying the end of PBH heating until well after the end of recombination.


In \ref{fig:boost_factor} we plot the ionisation boost factor, $\Delta x_\mathrm{H}$, as a function of redshift for each of the eleven PBH simulations. The solid and dotted coloured lines show the simulations with an $M^{-1}$ and $M^{-2}$ IMF, respectively. As expected, increasing the density of PBH in the simulations also causes $\Delta x_\mathrm{H}$ to increase. At the highest PBH densities ($10^{-2}~\Omega_C$), $\Delta x_\mathrm{H}$ saturates at unity. This indicates that PBH densities above $10^{-2}~\Omega_C$ would fully ionise their surrounding environment for an extended period.

For a fixed density of PBHs, changing the IMF shape from $M^{-1}$ to $M^{-2}$ increases the boost factor substantially (by a factor of 6.4 on average between redshifts $z=5000$--1000). This occurs because simulations with a $M^{-2}$ IMF are biased towards lighter PBHs, which evolve more rapidly than their heavier counterparts. The energy release from evaporating PBHs follows a highly non-linear temporal profile, with the 
majority of energy being ejected during the brief period immediately preceding total evaporation \citep{Auffinger2023}. Consequently, at any given redshift, simulations featuring more low-mass PBHs will contain a greater population of highly evolved PBHs approaching the end of their lifespans. This results in the $M^{-2}$ IMF producing substantially more PBH heating than the $M^{-1}$ IMF at equivalent redshifts. This can be seen in \ref{fig:boost_factor} by comparing PBH1 (solid-blue) and PBH6 (dotted-blue). 

The simulations with a higher $M_\mathrm{max}$ (PBH9-11) result in significantly larger $\Delta x_\mathrm{H}$ peaks, than simulations with the same $f_\mathrm{C,PBH}$, and have these peaks shifted to much later in time ($z<500$ instead of around). PBH11 with a density 100 times less than PBH5 both fully saturate at unity, however, PBH11 does so for a longer period of time. The more massive PBHs in PBH9-11 result in significantly more energy being released into the Universe, but at a much later time.

This lack of cooling has important implications for our redshift measurements of the matter-radiation decoupling. The instantaneous ceasing of heating in our simulations represents an idealised scenario that produces a sharp cutoff in $\Delta x_\mathrm{H}$. However, a more realistic treatment incorporating proper cooling mechanisms would distribute the decline in heating over an extended period, creating a gradual tapering rather than an abrupt end. Consequently, the redshift at which matter and radiation decouple (the redshift of the CMB) would be shifted to lower values, as the matter should maintain higher temperatures for a longer duration. Our measured redshift values for matter-radiation decoupling represent lower limits because our simulations and analysis can only capture the instantaneous cut-off point. The true redshift at which PBH heating becomes negligible would occur at lower redshifts than our results will indicate, with a larger difference occurring in the simulations with the largest heating.

\subsection{Gray body factors}
 GBFs quantify the frequency dependent transmission probabilities of particles escaping the PBH’s gravitational potential. They
modulate the otherwise blackbody Hawking spectrum
and are essential for accurate predictions of observational signatures.  To address this, \cite{yuan2026} 
derive analytic approximations for numerically
computed GBFs, using the recently developed \texttt{GrayHawk}
package M. 
\citep{calza} 
. For a Schwarzschild PBH of 2 $\times$ 10$^{-17}$ M$_\odot$ they find $\epsilon~\approx$ 2\% 
best approximate the \texttt{GrayHawk} results. For PBHs, unlike stellar mass black holes, there is no reason to expect spin up from conservation of initial angular momentum, but spin may be acquired later from neighbouring halos during galaxy formation.
Charged PBH may be a possibility, however \citep{p77}. 

\section{Results} \label{sec:Results}
\subsection{Effect on recombination} \label{sec:effect on recombination}
\begin{figure}
    \includegraphics[width=\linewidth]{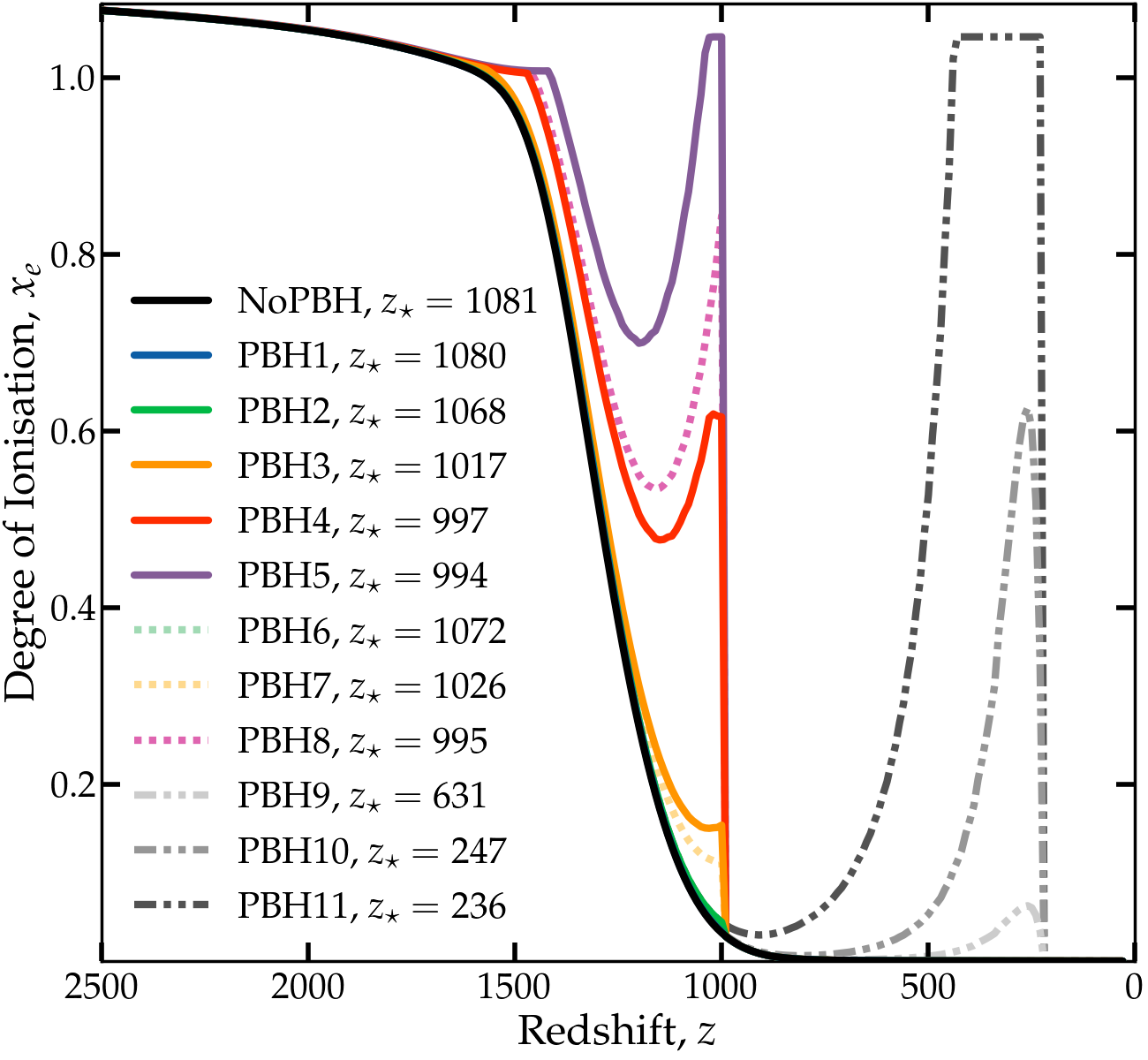}
    \caption{The \texttt{recfast} recombination history with (coloured) and without (black) heating due to evaporating PBH as a function of redshift. The vertical axis, $x_e$, is the degree of ionization, or the number of free electrons per baryon. The $z_\star$ in the legend is redshift of photon decoupling (i.e. the surface of last scattering). The delay in surface of last scattering is directly due to the ionising radiation from evaporating PBHs. 
    Simulation PBH7 (highest density of PBH) is fully ionised all the way down to redshift $z=0$. Indicating that with this density of PBHs or higher, the Universe would never become neutral as the PBHs are constantly ionising the gas.}
    \label{fig:x_e}
\end{figure}
The degree of ionisation, $x_e$, is the number of free electrons per baryon. If the Universe was fully ionised Hydrogen, then $x_e=1$. When Helium is partially or fully ionised, $x_e$ can exceed unity. For a fully ionised plasma of primordial composition (75\% H, 25\% He), $x_e \approx 1.08$. During recombination, $x_e$ drops rapidly as the Universe cools due to its adiabatic expansion. This rapid decline in $x_e$, allows the Universe to be transparent to photons, and, if the density of electrons is low enough, marks the redshift of the surface of last scattering.

Using the ionisation history, we can compute the surface of last scattering redshift, $z_\star$, from the optical depth, $\tau(z)$. This is the redshift at which matter and radiation decouple, and the Universe becomes `transparent'. We compute $z_\star$ by finding the lowest redshift such that $\tau(z) \geq 1$ where
\begin{equation}
   \tau(z) = \int_0^z n_e(z')\sigma_\mathrm{T}~c\frac{\mathrm{d}t}{\mathrm{d}z'}~\mathrm{d}z'\,.
\end{equation}
Here $\mathrm{d}t/\mathrm{d}z$ is calculated from assuming the scaling relation $t(z) = t_\mathrm{eq}(1+z)^{-3/2}$ for a matter dominated Universe ($z < 3400$), with the time of matter-radiation equality being $t_\mathrm{eq}\approx 1.48\times 10^{12}~\mathrm{s}$. The number density of electrons $n_e(z)$ is determined from the degree of ionisation as follows
\begin{equation}
    n_e(z) = x_e(z) \frac{\Omega_b \rho_c (1+z)^3}{\mu m_p}\,,
\end{equation}
with $\rho_c$ being the critical density at redshift $z=0$.

In \ref{fig:x_e}, we plot $x_e$ from the nine \texttt{recfast} recombination simulations, with the solid black line indicating the simulation without PBH and the coloured lines the simulations with PBH heating. The standard $\Lambda$CDM model without PBHs (NoPBH) shows a smooth transition from fully ionised to neutral around $z\approx 1100$, with the matter-radiation decoupling occurring at $z_\star = 1081$. As the fraction of PBHs increases, we observe a systematic delay in recombination, with the surface of last scattering occurring at progressively lower redshifts: $z_\star = 1080, 1068, 1017$ for PBH1 ($f_\mathrm{C,PBH} = 10^{-5}$), PBH2 ($f_\mathrm{C,PBH} = 10^{-4}$) and PBH3 ($f_\mathrm{C,PBH} = 10^{-3}$) respectively. 

 Simulations PBH4 and PBH5 with higher PBH fractions ($f_\mathrm{C,PBH} = 10^{-2.5}$ and $f_\mathrm{C,PBH} = 10^{-2}$ respectively) exhibit significant secondary ionisation peaks after initial recombination. PBH4 reaches a secondary peak of $x_e \approx 0.62$ at $z \approx 1000$, while PBH5 exhibits a larger secondary peak, with $x_e$ returning to unity at $z \approx 1000$. These secondary peaks would produce substantial alterations to the CMB power spectrum that are not observed. We conclude that for a $M^{-1}$ IMF, a PBH density of $\Omega_\mathrm{PBH} = 10^{-2}~\Omega_C$ must be an absolute upper limit. Above this value, $x_e$ would be at or above unity for a substantial period of time longer than $\Lambda$CDM models.

Switching to an $M^{-2}$ IMF results in substantially higher degrees of ionisation compared to the $M^{-1}$ IMF at equivalent PBH densities. For a $M^{-2}$ IMF, there is a larger proportion of PBHs of lower mass that release ionising radiation earlier in cosmic history, and this is clearly seen in \ref{fig:boost_factor} for PBH7 and PBH8. The $M^{-2}$ IMF produces ionisation boost factors approximately 6.4 times larger on average (between redshifts $z=5000$--1000) than the $M^{-1}$ IMF at the same PBH density. Consequently, even modest PBH densities of $\Omega_\mathrm{PBH} = 10^{-4}~\Omega_C$ with an $M^{-2}$ IMF (PBH7) produce ionisation levels comparable to much higher densities ($\Omega_\mathrm{PBH} = 10^{-3}~\Omega_C$) with an $M^{-1}$ IMF (PBH3).

The three simulations with higher maximum masses (PBH9, PBH10, and PBH11, where $M_\mathrm{max} = 10^{-17.5}~\mathrm{M}_\odot$) exhibit dramatically different recombination histories compared to their lower $M_\mathrm{max}$ counterparts. Because these simulations contain more massive PBHs with longer evaporation timescales, the bulk of their energy injection occurs significantly later in cosmic history. As shown in \ref{fig:boost_factor}, the ionisation boost factors for PBH9--11 peak at much lower redshifts ($z \lesssim 500$) compared to PBH1--8, which peak around $z \sim 1000$--1500.

As a consequence, in PBH9, PBH10, and PBH11, the surface of last scattering would no longer occur at high redshifts ($z \sim 1100$) as observed in the CMB, but instead would be delayed to $z \lesssim 600$. For PBH10 and PBH11, the surface of last scattering would occur at extremely low redshifts ($z_\star = 248$ and $236$ respectively), shifting the CMB from the expected epoch by more than 800, which would be completely inconsistent with Planck observations. Such dramatic delays would be observable in the \cite{Planck2020} observations of the CMB power spectrum. Additionally, the continued ionisation at $z < 100$ from these massive PBHs would be incompatible with IGM ionisation measurements at low redshifts \citep{saha25}.  This is investigated further in the Appendix.

We therefore infer that PBHs with an $M^{-2}$ IMF (or any IMF favouring lower-mass PBHs) extended beyond $10^{17.5}~\mathrm{M}_\odot$ would lead to an ionisation history that is fundamentally incompatible with CMB observations, as the heating would occur far too late in cosmic history.

In Figure \ref{fig:x_e}, all PBH models exhibit sharp cutoffs in their ionisation curves, most easily seen in PBH4, PBH5, PBH10 and PBH11. 
These abrupt transitions occur when the simulated population of PBHs have completely evaporated, terminating the additional energy injection into the surrounding gas. As described in more detail in \ref{sec:simparams}, this causes our measurements of $z_\star$ to be lower limits.

\subsection{Effect on the Hubble constant} \label{sec:Hubble_Constant}
\begin{table}
    \centering
    \begin{tabular}{cccclc}

         \hline
         Name & $f_\mathrm{C,PBH}$ & $z_\star$ & $\Delta z$& $\Delta \mathrm{H_0} / \mathrm{H_0}$& Plot Colour  \\
         \hline
         NoPBH & 0.00 & 1081 & $-$ & - & \textcolor{NoPBHColor}{$\blacksquare$} \\
         PBH1 & $10^{-5}$ & 1080 & 1 & $>$0.1\% &\textcolor{PBH1Color}{$\blacksquare$} \\
         PBH2 & $10^{-4}$ &  1068 & 13 & $>$1.8\% &\textcolor{PBH2Color}{$\blacksquare$} \\
         PBH3 & $10^{-3}$ &  1017 & 64 & $>$8.9\% &\textcolor{PBH3Color}{$\blacksquare$} \\
         PBH4 & $10^{-2.5}$ &  997 & 84 & $>$12.6\% &\textcolor{PBH4Color}{$\blacksquare$} \\
         PBH5 & $10^{-2}$ &  994 & 87 & $>$13.1\% &\textcolor{PBH5Color}{$\blacksquare$} \\
         PBH6 & $10^{-5}$ &  1072 & 9 & $>$1.2\% &\textcolor{PBH6Color}{$\blacksquare$} \\
         PBH7 & $10^{-4}$ &  1026 & 73 & $>$8.0\% &\textcolor{PBH7Color}{$\blacksquare$} \\
         PBH8 & $10^{-3}$ & 995 & 88 & $>$12.9\% & \textcolor{PBH8Color}{$\blacksquare$}\\
         PBH9 & $10^{-6}$ &  631 & 450 & $>$88\% &\textcolor{PBH9Color}{$\blacksquare$} \\
         PBH10 & $10^{-5}$ &  248 & 833 & $>>$100\% &\textcolor{PBH10Color}{$\blacksquare$} \\
         PBH11 & $10^{-4}$ & 236 & 845 & $>>$100\% & \textcolor{PBH11Color}{$\blacksquare$}\\

         \hline
    \end{tabular}
    \caption{The results of the \texttt{recfast} simulations. From left to right, the column names are: simulation name, the fraction of dark matter in the form of PBH ($f_\mathrm{C,PBH}$), the redshift of matter-radiation decoupling ($z_\star$), the change in decoupling redshift relative to NoPBH ($\Delta z$) the fractional change in the Hubble constant ($\Delta \mathrm{H_0}/\mathrm{H_0}$), and colour on the plots.}
    \label{tab:simsresults}
\end{table}

As shown in \ref{sec:effect on recombination}, evaporating PBHs cause additional heating and ionisation in the early Universe, which results in surface of scattering being shifted towards lower redshifts. This shift in the surface of last scattering also changes the value of $\mathrm{H}_0$. 

We quantify the effect PBHs have on the Hubble constant by the percentage change,
\begin{equation}
    \frac{\Delta \mathrm{H_0}}{\mathrm{H_0}} = \frac{3}{2} \Delta z (1 + z_\star)^{-1}\,,
    \label{eq:delta_H0}
\end{equation}
where $\Delta z$ is the change in $z_\star$ between the PBH simulation and the NoPBH simulation. See \ref{app:H0_H}1 for the full derivation of this equation.

\begin{figure}
    \centering
    \includegraphics[width=\linewidth]{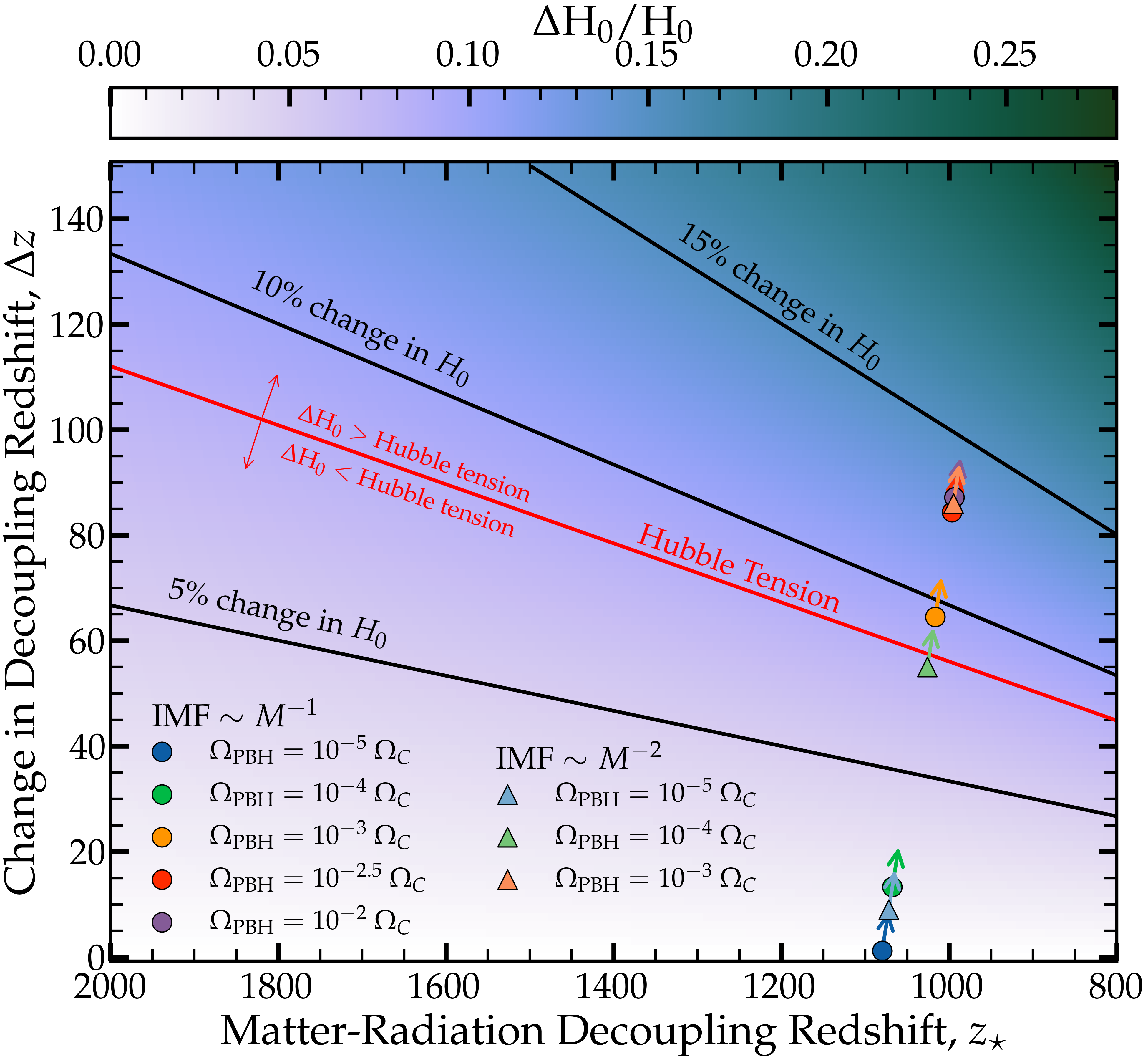}
    \caption{The change in $\mathrm{H_0}$ as a function of $\Delta z$ and the matter-radiation decoupling redshift, $z_\star$. The solid black lines indicate a 5\%, 10\% and 15\% increase in $\mathrm{H_0}$. The solid red line is the 8.7\% change that would be required to shift the 67.37 $\mathrm{km~s^{-1}~Mpc^{-1}}$ value from \cite{Planck2020}, to the most constrained local distance ladder measurement of 73.2 $\mathrm{km~s^{-1}~Mpc^{-1}}$ from \cite{Riess2022}. The eight scatter points are $\Delta \mathrm{H_0}/\mathrm{H_0}$ from PBH1-8, with circles and triangles representing the $M^{-1}$ and $M^{-2}$ IMF simulations respectively. The arrows indicate the direction in which the points would shift if a more comprehensive cooling scheme were included in \texttt{recfast}. A cosmic PBH density of $\Omega_\mathrm{PBH} \approx 10^{-4} - 10^{-3}~\Omega_C$ (simulations PBH3 and PBH7), would be enough to alleviate the Hubble tension entirely. However, a smaller amount of PBH would still cause a significant change in $\mathrm{H_0}$.}
    \label{fig:dH0onH0}
\end{figure}

In \ref{tab:simsresults}, we list the value of $z_\star$, $\Delta z$ and $\Delta \mathrm{H_0}/\mathrm{H_0}$ for each of the simulations.
In \ref{fig:dH0onH0}, we show how the matter-radiation decoupling redshift ($z_\star$) and the change in decoupling redshift ($\Delta z$) caused by PBH heating for our different simulation models affect the value of $\mathrm{H}_0$. The color gradient represents the relative change in the Hubble constant ($\Delta \mathrm{H}_0/\mathrm{H}_0$), with darker blue regions indicating larger changes up to approximately 30\%. The solid black lines denote a 5\%, 10\%, and 15\% change in $\mathrm{H}_0$ and the colored scatter points are the results from the PBH simulations with an $M^{-1}$ IMF.

As we described earlier, the estimates on $\Delta \mathrm{H}_0$ due to PBH are lower bounds. We use arrows to indicate the direction the data point would move if a more comprehensive cooling scheme were included in \texttt{recfast}.



If the entirety of the Hubble tension is caused by inaccurate modelling of the CMB and recombination due to PBH evaporation, then the initial cosmic energy density of PBHs, $\Omega_\mathrm{PBH}$, would be between $10^{-4}$ and $10^{-3}~\Omega_C$. A $M^{-2}$ IMF, required a lower density of PBHs to change the sound horizon the same amount as a $M^{-1}$ IMF. 

That said, it is extremely unlikely that the entirety of the Hubble tension is due to PBH. \cite{Jedamzik2021} showed that reducing the redshift of the sound horizon alone will introduce other tensions in cosmology, namely baryonic acoustic oscillations (BAO) \citep{Alam2017} and galaxy weak lensing \citep{Asgari2021, Abbott2018}. However, a small portion of the Hubble tension may be caused by recombination reheating due to a low density of PBHs in the early Universe. Even a $10^{-4}~\Omega_C$ can change the value of $\mathrm{H}_0$ by approximately 2\%, which is larger than the uncertainty  in $\mathrm{H}_0$ as measured by SH0ES \citep{Riess2022} and Planck \citep{Planck2020} (1.4\% and 0.74\% respectively).  

\subsection{Constraints on PBH masses and abundances}
\subsubsection{Extragalactic gamma-ray background}
PBHs in the range 10$^{13}$-10$^{18}$ g that survive to low redshift contribute to 
the isotropic diffuse $\gamma$-ray background. Constraints from COMPTEL
and Fermi-LAT on this background 
restrict f$_{C,PBH}$. 
PBH masses in their phase of major mass loss (their half mass time) at
10 $<$ z $<$ 0 contributing to the diffuse $\gamma$-ray background are approximately
10$^{-18}$ to 10$^{-17}$ M$_\odot$ (2 $\times$ 10$^{-15}$ to 2 $\times$ 10$^{-16}$ g). The mass range
that we are considering here, evaporating during and before recombination
at 4000 $<$ z $<$  1000, is 10$^{-20}$ to 10$^{-19}$ M$_\odot$.  
The constraint found by \cite{czl}
is f$_{PBH}~<$ 10$^{-5}$ at 10$^{13.5}$ g,
falling to f$_{PBH}~ <$ 10$^{-7.5}$ at 10$^{14.8}$ g. 
However, at the current epoch, z = 0, the Hawking radiation they consider, peaking at 2 MeV for 1.75 $\times$ 10$^{-17}$ M$_\odot$ and 70 MeV for 5 $\times$ 10$^{-19}$ M$_\odot$, has redshifted to 2 keV and 70 keV respectively, outside the range of the $\gamma$-ray telescopes.

\subsubsection{Positron annihilation and the 511 keV line}
For PBHs with masses M $\sim$ 10$^{15}$-10$^{16}$ g, the Hawking temperature T$_H~\sim$  m$_ec^2$ and
copious electron-positron pair production results in a 511 keV
annihilation signal. INTEGRAL/SPI observations of this line -- in particular,
the detection toward the Large Magellanic Cloud, which provides a
constraint freer of Galactic astrophysical uncertainties,
place strong bounds on PBHs in this mass range \citep{kp23}. 
Simulations PBH9-11, which use M$_{max}$ = 10$^{-17.5}$ M$_\odot$ to 6 $\times$ 10$^{15}$ g, contain
PBHs that survive well past recombination and fall within this constrained
regime.

There is a considerable literature on annihilation radiation from the
Galactic Center found by Fermi-LAT and confirmed by INTEGRAL.
It is the best evidence available that a major component of the dark matter is
WIMPs that decay through various channels leading to 511 keV emission.
However, there is also
the possibility that the signal originates in millisecond pulsars
from high temperature accretion disks. Hawking radiation can only
be subdominant in the region of interest, because the spectrum does
not show the GeV peak expected from PBHs evaporating in the last few Gyrs
(Mould \& List in preparation). The findings in the present paper
are consistent with the f$_{PBH}~<$ 10$^{-3}$ constraint of \cite{laha} 
in the 10$^{16}$ to 10$^{17}$ g range. 

\subsubsection{Voyager 1 electron and positron flux}
In-situ measurements of the low-energy cosmic-ray electron and
positron spectrum by Voyager 1 constrain the local rate of PBH evaporation.
\cite{bc2019}  
find that
PBHs with a mass M $\leq$ 10$^{17}$ g are expected to
inject sub-GeV electrons and positrons in the Galaxy via Hawking radiation.
These cosmic rays are shielded by the solar magnetic field for Earth-bound
detectors, but not for Voyager 1, which is now beyond the heliopause. 
They use its data to constrain the fraction of PBHs to the dark matter in the
Galaxy, finding that PBHs with M $<$ 10$^{16}$ g cannot contribute more than
0.1\% (or less for a log-normal mass distribution).
\begin{figure}
    \centering
    \includegraphics[width=\linewidth]{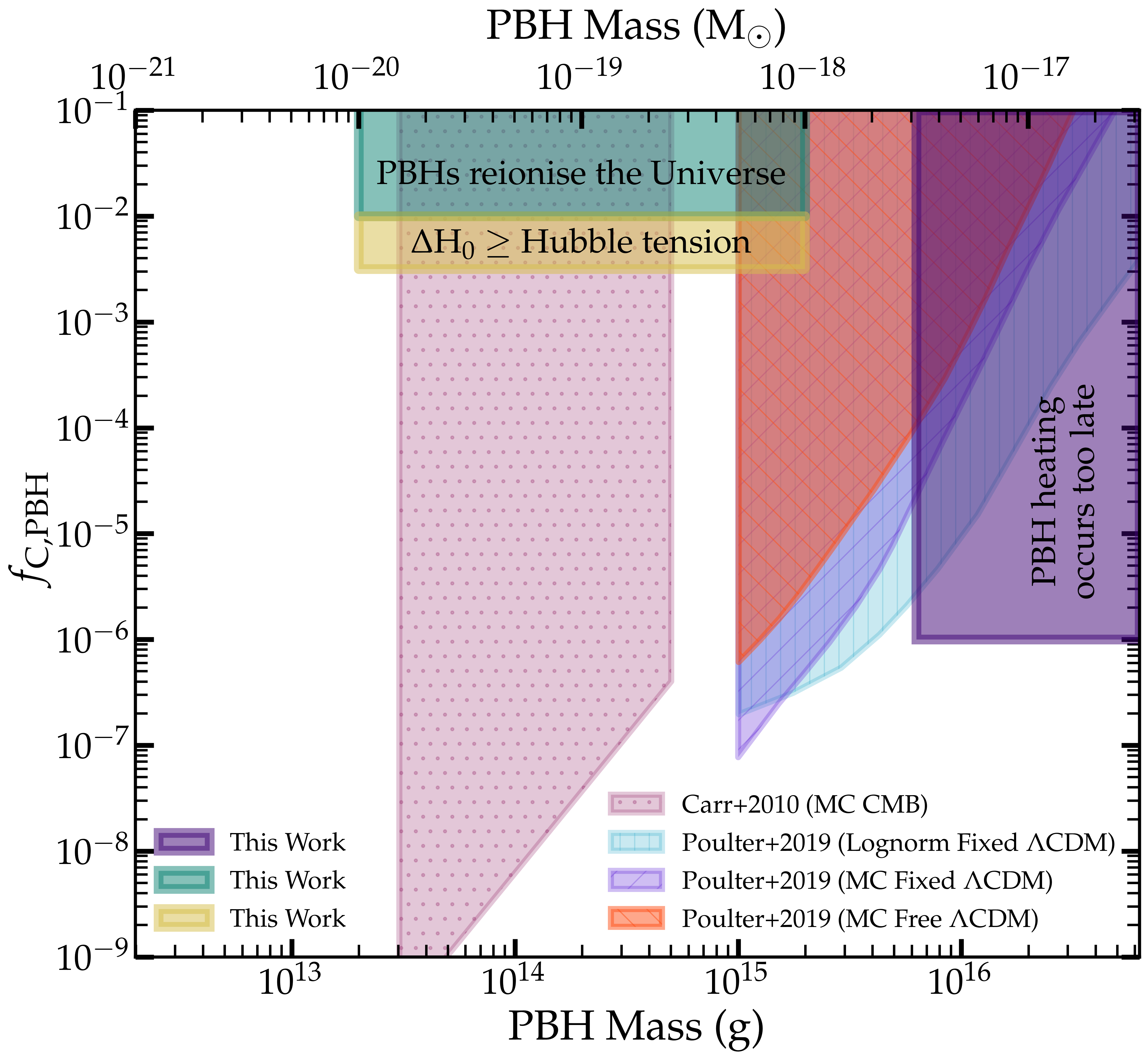}
    \caption{The exclusion regions for PBH masses and densities. In green we have excluded densities of $f_{\mathrm{C,PBH}} \geq 10^{-2}$, as above this, the additional heating from PBHs causes the Universe to become reionised shortly after recombination. In yellow, assuming an IMF distribution of $M^{-1}$, we have excluded the region that results in PBH heating, reducing the CMB redshift such that the CMB estimate of $\mathrm{H}_0$ would be larger than the local measurements. In purple, we have excluded PBHs with masses above $10^{-17.5}~\mathrm{M_\odot}$, as these cause reheating to occur too late into the thermal history of the Universe. We have also overplotted the constraints from the works of \cite{Carr2010} and \cite{Poulter2019}. Here, `MC' in the legend indicates that these are \emph{monochromatic} constraints, which are complemented by our \emph{non-monochromatic} results.
    }
    \label{fig:exclusions}
\end{figure}

In Figure \ref{fig:exclusions} we plot the exclusion regions we have made on PBH masses and densities using our \texttt{recfast} simulations.

Firstly, as mentioned in \ref{sec:effect on recombination}, for an IMF of $M^{-1}$, a PBH density of $\Omega_\mathrm{PBH} = 10^{-2}~\Omega_C$ must be an absolute upper limit in the mass range $10^{-20}$ to $10^{-18}~\mathrm{M_\odot}$. For PBH fractions above this value, the Universe would become reionised shortly after recombination ($x_e$ at unity). Changing to a steeper $M^{-2}$ IMF increases the average ionisation strength by a factor of 6.4. As a consequence, the absolute upper limit would also hold for $M^{-2}$ IMFs (and presumably for steeper IMFs as well). We indicate this constraint with a green box in \ref{fig:exclusions}, labeled `PBHs reionise the Universe'.

Secondly, we use the Hubble tension as a constraint on PBHs. Values of $f_\mathrm{C, PBH}$ between $10^{-2.5}$ and $10^{-2}$ do not fully reionise the Universe, but would change the value of $H_0$ enough that the early-time measurement (CMB) would become larger than the late-time measurement (local distance ladder). These values would cause an `inverted' Hubble tension. We indicate this constraint in \ref{fig:exclusions} with the yellow box labeled with `$\Delta H_0 \geq$ Hubble tension'. Based on \ref{fig:dH0onH0}, we could extend this constraint down to $10^{-3}$, as the simulation PBH3 has a $\Delta H_0$ larger than the Hubble tension (8.7\%). 
However, we have chosen to take the more conservative constraint of $10^{-2.5}$. The reason for choosing $10^{-2.5}$ is that PBH3 is extremely close to the Hubble tension line in \ref{fig:dH0onH0}, and simply changing the IMF shape (a power law with an index less than $-1$) can move this value towards (power law with an index greater than $-1$) or away from the break point.  Further support for this limit is in Appendix A4.

Thirdly, PBHs with masses $> 10^{-17.5}~\mathrm{M}_\odot$ would evaporate and inject ionising energy at redshifts $z < 100$. This late-time ionisation can be ruled out by observations of the intergalactic medium at $z \sim 10$ using the Lyman-$\alpha$ forest \citep{saha25}, which constrain the IGM temperature and ionisation state during this epoch. These observations demonstrate that the ionisation fraction remains low at these redshifts, incompatible with significant PBH evaporation occurring at redshifts $z < 100$. We indicate this constraint in \ref{fig:exclusions} with a purple box labelled `PBH heating occurs too late'.

In \ref{fig:exclusions}, we also include the constraints from the works of \cite{Carr2010} and \cite{Poulter2019}, which are in the same mass ranges we are considering. The \cite{Carr2010} constraints are on monochromatic PBHs using the CMB E-Mode polarisation autocorrelation ($\mathcal{D}_l^{EE}$ or EE) and the temperature-E-mode polarisation cross-correlation ($\mathcal{D}_l^{TE}$ or TE). Because \cite{Carr2010} used a monochromatic model, these are not 1-to-1 comparisons with our non-monochromatic simulations with power-law mass distributions. As mentioned earlier, using non-monochromatic PBH distributions allows for a higher fraction of PBHs under the same constraints.

We have included three results from \cite{Poulter2019}: two monochromatic results and one non-monochromatic. The non-monochromatic result (Lognorm Fixed $\Lambda$CDM) is the most similar to our work, as it uses a mass distribution (a log-normal distribution with $\sigma=10$) with fixed $\Lambda$CDM parameters. Their constraint of $f_\mathrm{C,PBH} < 10^{-4}$ for this log-normal distribution is comparable to our $M^{-1}$ IMF results, though the different mass distribution shapes make direct comparison challenging. The monochromatic constraints are significantly tighter, as expected, because concentrating all PBH mass at a single value produces stronger observable effects than spreading it across a mass range.

\cite{Poulter2019} showed that by allowing $\Lambda$CDM parameters to vary simultaneously with PBH fraction and mass when fitting the CMB power spectrum (MC Free $\Lambda$CDM, hatched-red in \ref{fig:exclusions}), the constraints on $f_\mathrm{C,PBH}$ are relaxed by an order of magnitude, compared to the fixed $\Lambda$CDM case (MC Fixed $\Lambda$CDM, hatched-lilac in \ref{fig:exclusions}). For a uniform mass distribution of PBH between $5\times10^{-19}$ and $5\times 10^{-17}$ $\mathrm{M}_\odot$ (not shown on \ref{fig:exclusions}) they find $f_\mathrm{C,PBH}< 1.6\times10^{-5}$ when allowing $\Lambda$CDM parameters to vary.

Overall, our results are consistent with \cite{Carr2010} and \cite{Poulter2019}, and have extended the exclusion regions. We again remind the reader that these $f_\mathrm{C,PBH}$ values are for primordial abundances. Over time $f_\mathrm{C,PBH}$ values decrease as PBHs evaporate, and at any redshift there is mass lower-limit, $M_\mathrm{PBH}$, for which all PBHs with lower masses have completely evaporated.

\subsection{Effect on the CMB Power Spectrum}
Having established that PBH fractions of $f_\mathrm{C,PBH}\approx10^{-3}$ could shift the sound horizon significantly, we now investigate whether such densities would produce observable signatures in the CMB power spectrum that would have been detected by Planck. To determine if the density of PBHs we use in our simulations would be detectable in the CMB, we ran \texttt{CAMB} (Code for Anisotropies in the Microwave Background) \citep{LC2000} with our ionisation-redshift ($x_e$-$z$) relations obtained using \texttt{recfast}.

As described in \cite{Cheng25} and \cite{Zhang2007}, PBHs can affect the CMB power spectrum in several ways. Firstly, it is through the ionisation history. At large scales, polarisation and temperature anisotropies are generated by photon scattering due to free electrons (Thomson scattering). The presence of PBHs producing additional ionisation at high-redshifts would be visible at large scales (small $l$) in the polarisation autocorrelation (EE) and the temperature-polarisation cross-correlation (TE).

\begin{figure}
    \centering
    \includegraphics[width=\linewidth]{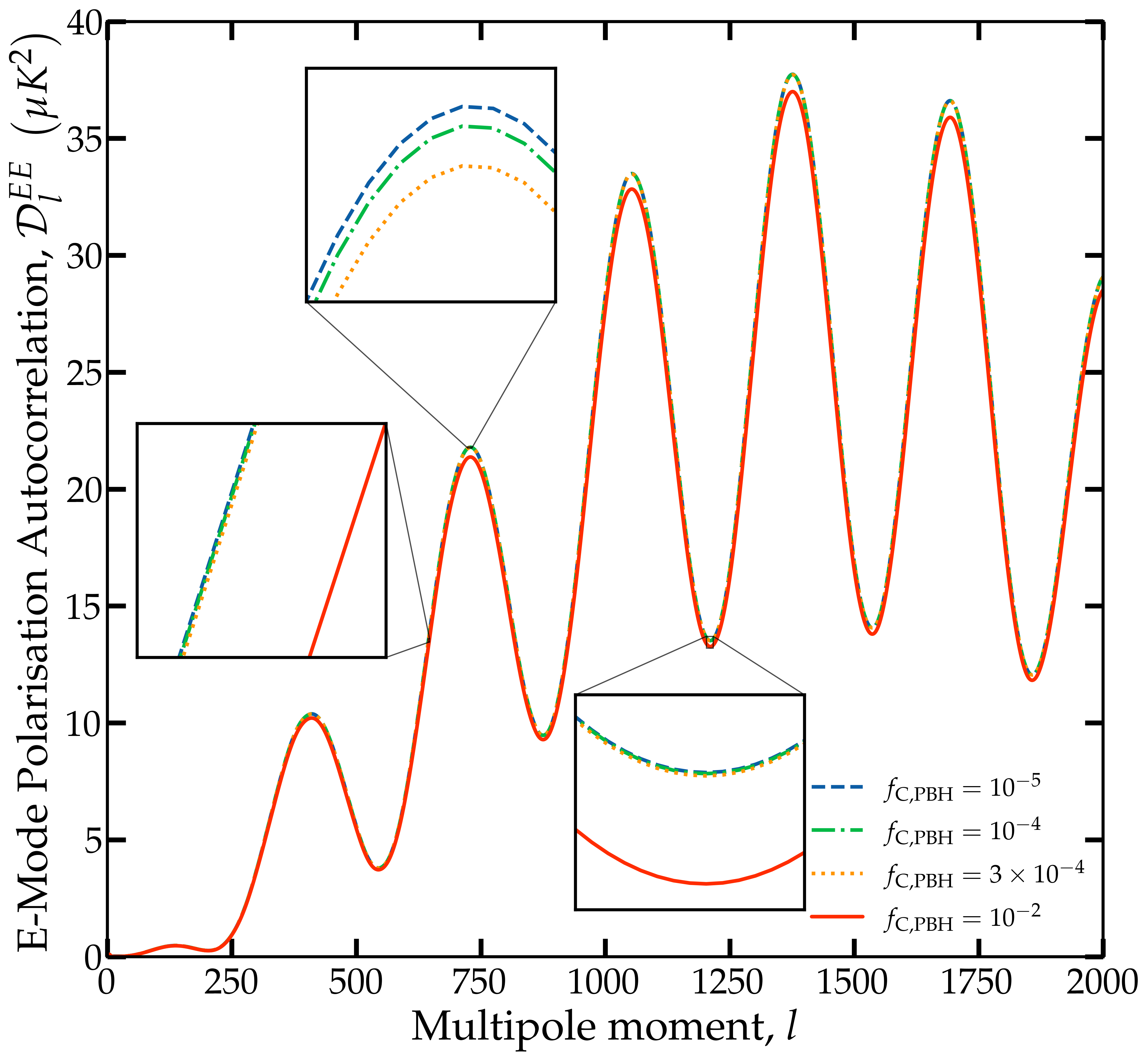}
    \caption{The \texttt{CAMB} E-Mode polarisation autocorrelation, $\mathcal{D}_l^{EE}$, simulations for various amounts of PBH fractions. The RMS difference between the lowest amount of PBH ($f_\mathrm{C,PBH}=10^{-5}$, blue-dashed) and the highest ($f_\mathrm{C,PBH}=10^{-2}$, red-solid) is 0.22 $\mu$K$^2$ over the range $0<l<2000$. This is less than 0.6\% of the peak signal value.}
    \label{fig:cambdata}
\end{figure}

In \ref{fig:cambdata}, we plot the EE autocorrelation for various PBH fractions between $f_\mathrm{C,PBH}=10^{-5}$ to $10^{-2}$. The solid-red line indicates the highest density of PBHs, with the dashed-blue line being the lowest.

There is only a slight difference in EE between the highest and lowest fraction of PBHs, most easily seen at the peaks at high $l$. The root mean square (RMS) difference in EE polarisation power in the range $0<l<2000$ is $0.22~\mu$K$^2$ between $f_\mathrm{C,PBH}=10^{-2}$ and $10^{-5}$. With a peak signal value of approximately $38~\mu$K$^2$, this means that there is only a 0.6\% difference in EE power between the maximum and minimum PBH simulations. 


\section{Conclusions} \label{sec:Conclusions}
In this paper, we have explored the effect that primordial black holes (PBHs) in the early Universe ($z>1100$) have on recombination, the Hubble constant, and as a consequence, the Hubble tension. If PBHs were evaporating during the era of recombination, they would cause additional heating in the early Universe that would delay the surface of last scattering and thereby the CMB. We performed recombination simulations that included the heating produced by evaporating PBHs in the mass range $10^{-20}$ to $10^{-17.5}$ M$_\odot$ ($\approx 10^{13} - 10^{16}$ g). 

We summarise our conclusions as follows:
\begin{itemize}
    \item PBHs in the mass range 10$^{-20}$--10$^{-18}$ M$_\odot$ are likely to have the most significant effect on recombination, as they entirely evaporate before redshift $z \sim 1100$ due to Hawking radiation.
    \item PBHs deliver most of their energy in the form of \emph{secondary ionisation}, heating from Compton scattering and pair-production losses rather than through \emph{direct ionisation}. This is because the photoelectric effect cross-section decreases rapidly with the high photon energies characteristic of Hawking radiation from PBHs.
    \item The change in temperature of the gas surrounding an entirely evaporated PBH is independent of the initial PBH mass. This temperature change is approximately $\Delta T_\mathrm{gas} = 2\times10^{13}~f_\mathrm{C,PBH}$ K, where $f_\mathrm{C,PBH}$ is the PBH fraction of dark matter. This independence of PBH mass on temperature change is due to more massive PBHs being surrounded by a proportionally greater amount of gas.
    \item A cosmic PBH energy density of $\Omega_\mathrm{PBH} \approx 10^{-3}~\Omega_C$ with a $M^{-1}$ initial mass function would cause an 8.9\% increase in the value of $\mathrm{H}_0$, enough to entirely resolve the current Hubble tension between the early-time (CMB) and late-time (distance ladder) measurements. This corresponds to delaying recombination by $\Delta z \sim 64$.
    \item The value of $\mathrm{H}_0$ is extremely sensitive to $\Omega_\mathrm{PBH}$. Even PBH densities as low as $\Omega_\mathrm{PBH} \approx 10^{-4}~\Omega_C$ produce $>$1.8\% changes in $\mathrm{H}_0$, larger than the current measurement uncertainties from both Planck \citep{Planck2020} and SH0ES \citep{Riess2022}.
    \item We establish an absolute upper limit of $\Omega_\mathrm{PBH} < 10^{-2}~\Omega_C$ for PBHs in the $10^{-20}$--$10^{-18}~\mathrm{M}_\odot$ mass range. Beyond this density, PBHs would cause complete reionisation of the Universe shortly after recombination, maintaining $x_e \approx 1$ for extended periods.
    \item $\Omega_{PBH}$ values in the last three bullet points should be multiplied by a grey body factor 
    \item PBHs of masses $> 10^{-17.5}~\mathrm{M_\odot}$ are strongly constrained as they evaporate and ionise the IGM at $z<100$.
   \item The shape of the PBH initial mass function significantly affects the ionisation history. An $M^{-2}$ IMF (favouring lower-mass PBHs) produces ionisation boost factors approximately 6.4 times larger than an $M^{-1}$ IMF at the same PBH density, because lower-mass PBHs evaporate earlier and more rapidly.
    \item The Hubble tension places constraints on the properties of PBHs. The presence of approximately 0.1\% dark matter in the form of PBHs, causes the sound horizon of the CMB to be shifted by enough that the early-time measurement of $\mathrm{H_0}$ becomes larger than the \cite{Riess2022} value.
    \item To make progress with the Hubble tension in the presence of dark matter ionizers with properties that are not fully known, it may be better to find the range of ionization hostories consistent with the CMB\citep{groupm}, and from them find the range of possible values of H$_0$.
\end{itemize}

This work shows that PBH evaporation represents a physically plausible mechanism for modifying the early Universe's ionisation history without invoking new physics. While it is unlikely that PBHs alone account for the entire Hubble tension (modifications to the sound horizon introduce other cosmological tensions with baryon acoustic oscillations \citep{Jedamzik2021}) they could contribute meaningfully to the discrepancy. 

Our results suggest that measuring $\mathrm{H}_0$ locally at $z \lesssim 1$ may be more robust than inferring it from the CMB until we can confidently rule out small populations of evaporating PBHs in the early Universe.

\section*{Acknowledgements}
We thank the John Templeton Foundation for the grant "Two standard models meet." The ARC Centre of Excellence for Dark Matter Particle Physics is funded by grant CE200100008. Thanks also to Marko Laine for the MCMC code and to Yin Zhe Ma and Matias Zaldarriaga for their comments on the paper.

\subsection*{Software}
\begin{itemize}
    \item \texttt{Astropy} \citep{Astropy2013,Astropy2018},
    \item \texttt{CMasher} \citep{CMasher2020, CMasher2024},
    \item \texttt{Matplotlib} \citep{matplotlib2007},
    \item \texttt{Numpy} \citep{numpy2011,numpy2020},
    \item \texttt{recfast} \citep{Seager1999, Seager2000, Scott2009},
    \item \texttt{CAMB} \citep{LC2000}
    
\end{itemize}

\section*{Data Availability}
All simulation data, modified \texttt{recfast code}, analysis scripts, and figure generation code are publicly available on Github at \url{https://github.com/abatten/PBH-recfast}. \\
The original \texttt{recfast} code can be obtained from \url{https://www.astro.ubc.ca/people/scott/recfast.html}. \\
CAMB is available from \url{https://lambda.gsfc.nasa.gov/}.


\bibliography{references.bib} 


\appendix

\section{Derivation of change in Hubble Constant} \label{app:H0_H}
In this appendix, we show the derivation of $\Delta\mathrm{H}_0/\mathrm{H}_0$ that we use in \ref{sec:Hubble_Constant}.

\subsection{Fractional change in H = fractional change in Hubble Constant}
In this section, we will show that at high redshift ($z\gg 100$), near the surface of last scattering the following relation is true.
\begin{equation}
    \frac{\mathrm{dH}_0}{\mathrm{H}_0} \approx \frac{\mathrm{dH}}{\mathrm{H}}
\end{equation}

First, we will start with the Hubble parameter, $\mathrm{H}(z)$, and the Friedmann equation  scaling with large redshift ($z$),
\begin{equation}
    \mathrm{H}(z) = \mathrm{H}_0 \Omega_m (1+z)^{3/2}\,,
    \label{eq:Hparam_scaling}
\end{equation}
where $\Omega_m$ is the cosmic matter density. By using implicit differentiation and applying the product rule, we obtain the following, 
\begin{equation}
    \mathrm{dH} = \frac{3}{2}\mathrm{H}_0 \Omega_m (1+z)^{1/2} \mathrm{d}z + \mathrm{dH}_0 \Omega_m (1+z)^{3/2}\,.
\end{equation}
By factoring the $\Omega_m (1+z)^{3/2}$ term and dividing both sides we get the following,
\begin{equation}
    \frac{\mathrm{dH}}{\Omega_m (1+z)^{3/2}} = \frac{3}{2}\mathrm{H}_0 (1+z)^{-1} + \mathrm{dH}_0\,.
\end{equation}
By multiplying by a factor of $\mathrm{H}$ as shown in \ref{eq:Hparam_scaling} we get the following,
\begin{equation}
    \mathrm{H}_0 \mathrm{dH} = \frac{3}{2}\mathrm{H}\mathrm{H}_0 (1+z)^{-1} +\mathrm{H}\mathrm{dH}_0\,,
\end{equation}
We now divide both sides by $\mathrm{H}$ and $\mathrm{H}_0$ and finally rearrange for $\mathrm{dH}_0/\mathrm{dH}$ to get the following:
\begin{equation}
    \frac{\mathrm{dH}_0}{\mathrm{H}_0} = \frac{\mathrm{dH}}{\mathrm{H}} - \frac{3}{2}(1+z)^{-1}\,.
\end{equation}
The factor $3/2 (1+z)^{-1}$ is extremely small at high redshift, resulting in a 1\% difference at $z=150$, and significantly less at redshifts approaching $z\sim 1000$. For this, we can choose to neglect this term and find that
\begin{equation}
    \frac{\mathrm{dH}_0}{\mathrm{H}_0} = \frac{\mathrm{dH}}{\mathrm{H}}
\end{equation}
Hence, to find the fractional change in $\mathrm{H_0}$, we can evaluate the fractional change in the Hubble parameter $\mathrm{H(z)}$ at the time of the surface of last scattering ($z = z_\star$).

\subsection{Change in Hubble Parameter}
In the previous section we showed that to measure the change in the Hubble constant, $\Delta \mathrm{H_0}/\mathrm{H_0}$, we only need to evaluate the change in the Hubble parameter at the surface of last scattering, $z_\star$.

In this section, we relate $\Delta z_\star$ to $\Delta$H , i.e.  
\begin{equation}
    \frac{\mathrm{dH}}{\mathrm{H}} = \frac{3}{2} \frac{\mathrm{d}z}{1+z}\,.
\end{equation}
In a matter-dominated universe ($10 < z < 3400$), the scale factor $a~\propto~t^{2/3}$, or in other words the following equation is true at the time of recombination,
\begin{equation}
1+z = t^{-2/3} k^{-1}\,,
\end{equation}
where $t$ is time and $k$ is approximately constant. By differentiating both sides with respect to time, we have the following, 
\begin{equation}
    \dot{z} = \frac{\mathrm{d}z}{\mathrm{d}t} = -\frac{2}{3} t^{-5/3} k^{-1}\,.
\end{equation}
Multiplying both sides by $a$ gives us the following,
\begin{equation}
    \frac{\dot{z}}{1+z} = -\frac{2}{3} t^{-1}\,.
\end{equation}
The quantity $\dot{z}/(1+z)$ is equivalent to $\dot{a}/a = \mathrm{H}$. By substituting $\mathrm{H}$ and implicit differentiation we obtain
\begin{equation}
    \mathrm{H} = -\frac{2}{3} t^{-1}\,.
\end{equation}
\begin{equation}
    \mathrm{dH} = \frac{2}{3} t^{-2} \mathrm{d}t
\end{equation}
Now, dividing both sides by $\mathrm{H} = -\frac{2}{3}t^{-1}$, 
\begin{equation}
    \frac{\mathrm{dH}}{\mathrm{H}} = - t^{-1}\mathrm{d}t\,. 
\end{equation}
Using the relation $x^{-1} \mathrm{d}x = \mathrm{dln}x$,
\begin{equation}
   \frac{\mathrm{dH}}{\mathrm{H}} = -\mathrm{dln} ~t
\end{equation}
We now use the relation $\mathrm{dln}t = -\frac{3}{2}\mathrm{dln}(1+z) = -\frac{3}{2}(1+z)^{-1}\mathrm{d}z$
\begin{equation}
    \frac{\mathrm{dH}}{\mathrm{H}} = \frac{3}{2} \frac{\mathrm{d}z}{1+z}
\end{equation}
Hence we now have the following equation for $\mathrm{dH}_0/{\mathrm{H}_0}$,
\begin{equation}
 \frac{\mathrm{dH}_0}{\mathrm{H}_0} = \frac{\mathrm{dH}}{\mathrm{H}} = \frac{3}{2}\mathrm{d}z (1+z)^{-1}\,.
\end{equation}
Evaluating this at the surface of last scattering and change infinitesimals to $\Delta$'s we have \ref{eq:delta_H0}
\begin{equation}
 \frac{\Delta \mathrm{H}_0}{\mathrm{H}_0} = \frac{3}{2}\Delta z (1+z_\star)^{-1}\,.
\end{equation}

\subsection{The acoustic angle}
 While we have shown in the above that changing the age of the Universe at the surface of last scattering also changes H$^{-1}$ by the same fractional amount, there are other issues in
 making adjustments to z$_\star$ Planck and ACT measurements of the CMB determine the acoustic angle $\theta_\star~=~r_s/D_A$. However, it is a simple matter to numerically differentiate $\theta_\star$, and we find that d($r_s/D_A$)/dh and d($r_s/D_A$)/d$\Omega_m$ 
 are approximately equal. 
 Therefore it is possible to null out an increase in h with a decrease in $\Omega_m$, as shown in Figure A1.
 \begin{figure}
     \includegraphics[width=0.55\textwidth]{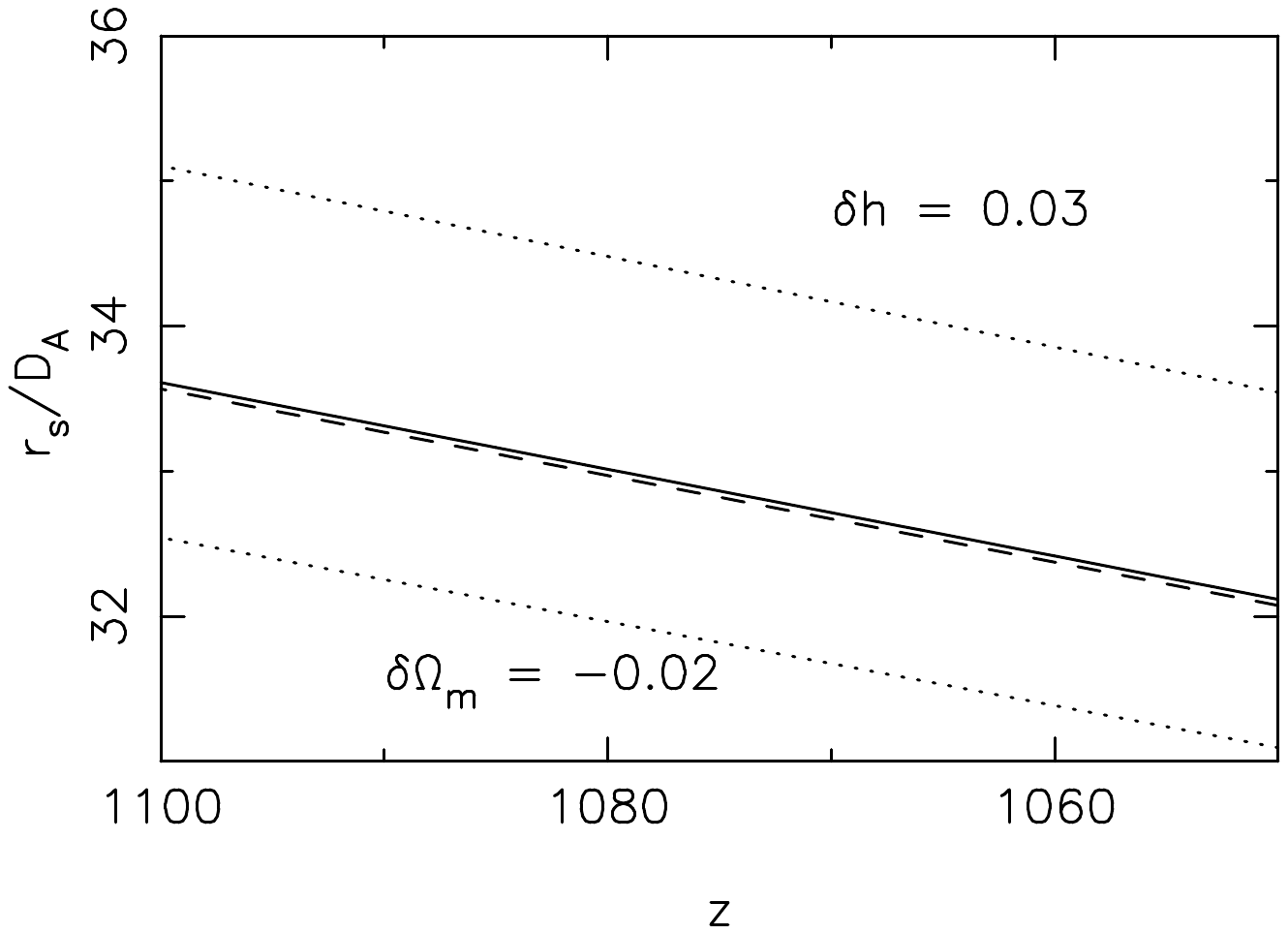}
     \caption{The acoustic angle $\theta*$ as a function of redshift (solid line: h = 0.67 and $\Omega_m$ = 0.3). The dotted lines are result of the indicated $\delta\Omega_m$ and $\delta h$ when z$_\star$ is moved by PBH to lower redshift. The dashed line illustrates nulling the two by $\delta\Omega_m$ = -0.024 and $\delta h$ = 0.036.}
 \end{figure}
 The dashed line in the figure amounts to $\delta \Omega_m h^2$ = 
 --0.015 (or a fractional decrease of 10\%).


\subsection{Markov Chain Monte Carlo analysis}
To further check for inconsistency with CMB measurements, MCMC was used to estimate H$_0$ and $\Omega_m$, using CAMB for 6 values of f$_{\rm PBH}$ treated as priors. For this we needed CMB data, and we used the latest Atacama Cosmology Telescope EE results \citep{ACT}. 

Their figure 40 shows 
that ACT has advanced the measurement of H$_0$ since the Planck mission. We used the CAMB model as a proxy for data with their H$_0$ (68.2) and $\Omega_m h^2$ (0.118) as the 2199 values of $l$ to fit. Additional priors from ACT were $\Omega_k$ = 0 and $\Omega_b h^2$ = 0.0226. The results are in Figure A2.
\begin{figure}
    \includegraphics[width=.65\columnwidth]{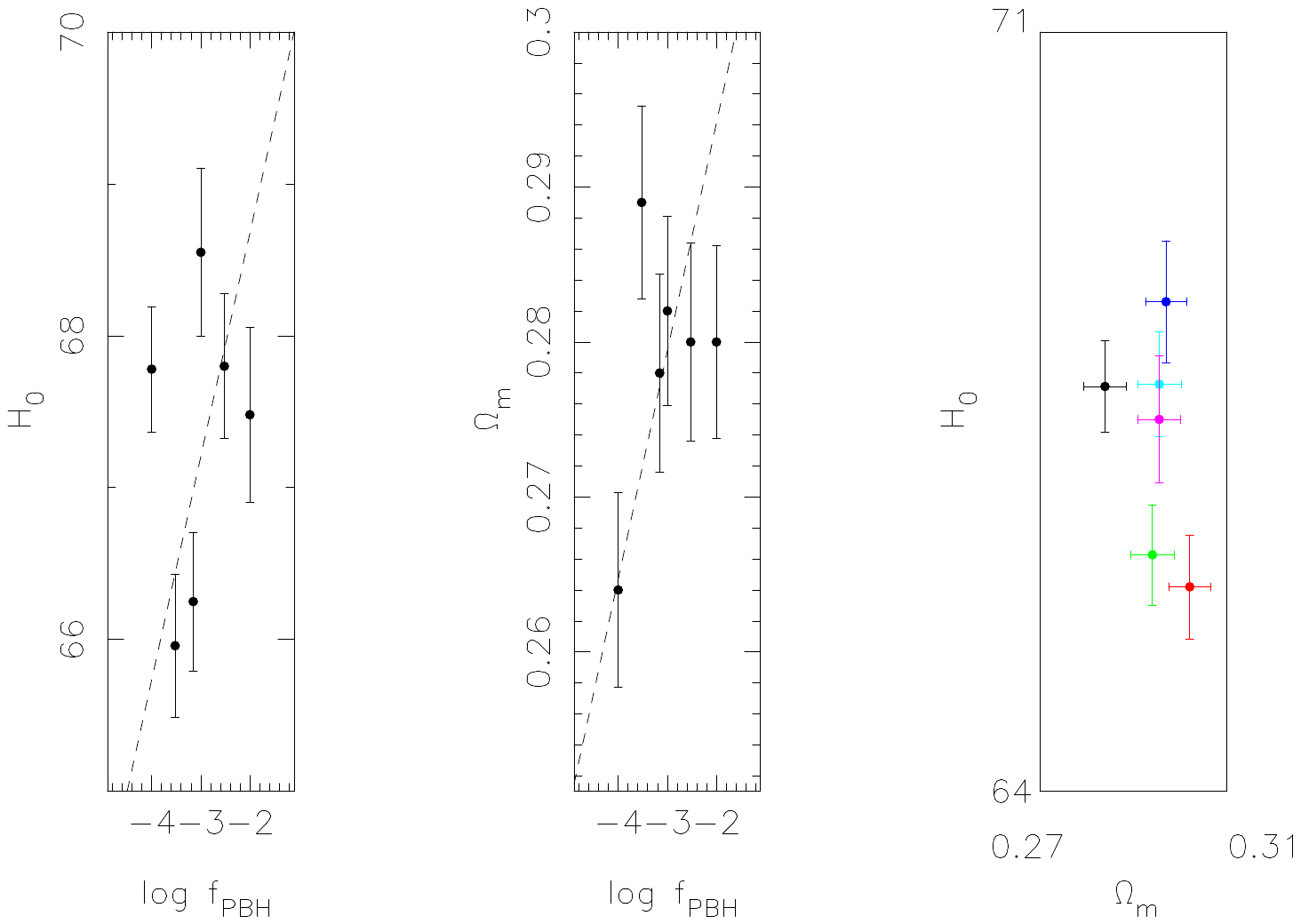}[h]
    \caption{Results of fitting six PBH ionization profiles to the CMB using MCMC. These include PBH 2, 3, 4 \& 5 in Table 2. Left and center: H$_0$ and $\Omega$ matter. right: the (H$_0$, $\Omega_m$) plane for these six values of f$_{\rm PBH}$. Red \& green are
    f = 3, 7 $\times$ 10$^{-4}$; blue and pink are f~$\ge$~10$^{-3}$. The dashed lines are an approximate fit.}
    \end{figure}
    
    This suggests that, while PBHs may achieve z$_*$ values raising H$_0$ by as much as 9\%, lower values are required to also fit the CMB. Other parameters in a CMB fit, the  scalar spectral index, n, the scalar amplitude, A$_s$, the optical depth, $\tau$, and, to a lesser extent, $\Omega_b$ can, for efficiency, be omitted from these MCMC runs, as they affect the amplitude of the acoustic oscillations, but not the positions of the peaks, leaving 
    only H$_0$ and $\Omega_m$.


\end{document}

%% file: references.bib
@ARTICLE{Gabadadze2024,
       author = {{Gabadadze}, Gregory and {Spergel}, David N. and {Tukhashvili}, Giorgi},
        title = {Inflation with the Trace Anomaly Action and Primordial Black Holes},
      journal = {arXiv e-prints},
         year = {2024},
          eid = {arXiv:2411.16834},
        pages = {arXiv:2411.16834},
          doi = {10.48550/arXiv.2411.16834},
archivePrefix = {arXiv},
       eprint = {2411.16834},
 primaryClass = {hep-th},
}

@book{Kohri2024, 
author = {Kohri, K.},
year = {2024},
title = {Primordial Black Holes},
publisher = {Springer Nature: Singapore},
pages = {440},
}

@article{saha25,
    author =  {Saha, A. and Kumar, S. and Singh, A. and  Parashari, P.
               and Laha, R.},
    title ={X} ,
    journal = {European Physical Journal C}, 
Volume ={85}, 
pages = {1117},
    year = {2025}
}

@article{H0DN,
       author = {{H0DN Collaboration} and {Casertano}, Stefano and {Anand}, Gagandeep and {Anderson}, Richard I. and {Beaton}, Rachael and {Bhardwaj}, Anupam and {Blakeslee}, John P. and {Boubel}, Paula and {Breuval}, Louise and {Brout}, Dillon and {Cantiello}, Michele and {Cruz Reyes}, Mauricio and {Cs{\"o}rnyei}, Geza and {de Jaeger}, Thomas and {Dhawan}, Suhail and {Di Valentino}, Eleonora and {Galbany}, Llu{\'\i}s and {Gil-Mar{\'\i}n}, H{\'e}ctor and {Graczyk}, Dariusz and {Huang}, Caroline and {Jensen}, Joseph B. and {Kervella}, Pierre and {Leibundgut}, Bruno and {Lengen}, Bastian and {Li}, Siyang and {Macri}, Lucas and {{\"O}z{\"u}lker}, Emre and {Pesce}, Dominic W. and {Riess}, Adam and {Romaniello}, Martino and {Said}, Khaled and {Sch{\"o}neberg}, Nils and {Scolnic}, Dan and {Sicignano}, Teresa and {Skowron}, Dorota M. and {Uddin}, Syed A. and {Verde}, Licia and {Nota}, Antonella},
        title = {The Local Distance Network: a community consensus report on the measurement of the Hubble constant at 1\% precision},
      journal = {arXiv e-prints},
         year = {2025},
        month = {oct},
          eid = {arXiv:2510.23823},
        pages = {arXiv:2510.23823},
archivePrefix = {arXiv},
       eprint = {2510.23823},
 primaryClass = {astro-ph.CO},
}

@article{LC2000,
    author = {Lewis, A. and Challinor, A. and Lasenby, A.},
    volume = {538},
    journal = {ApJ},
    title={Efficient Computation of Cosmic Microwave Background Anisotropies in Closed Friedmann-Robertson-Walker Models},
    year = {2000},
pages = {473}
}

@article{Cheng25,
author = {{Cheng}, Hanyu and {Yin}, Ziwen and {Di Valentino}, Eleonora and {Marsh}, David J.~E. and {Visinelli}, Luca},
title = {Constraining exotic high-z reionization histories with Gaussian processes and the cosmic microwave background},
journal = {arXiv e-prints},
year = {2025},
eid = {arXiv:2506.19096},
pages = {arXiv:2506.19096},
archivePrefix = {arXiv},
eprint = {2506.19096},
primaryClass = {astro-ph.CO},
}

@article{Sanchis,
author={Sanchis, H. and Barenboim, G. and Perez-Gonzalez, Y.},
journal={JCAP},
title={Efficient Computation of Cosmic Microwave Background Anisotropies in Closed Friedmann-Robertson-Walker Models},
volume={8},
pages={49},
year={2025},
}

@ARTICLE{Auffinger2023,
       author = {{Auffinger}, J{\'e}r{\'e}my},
        title = "{Primordial black hole constraints with Hawking radiation-A review}",
      journal = {Progress in Particle and Nuclear Physics},
         year = {2023},
        month = {jul},
       volume = {131},
          eid = {104040},
        pages = {104040},
archivePrefix = {arXiv},
       eprint = {2206.02672},
 primaryClass = {astro-ph.CO},
}

@ARTICLE{Musco2024,
       author = {{Musco}, Ilia and {Jedamzik}, Karsten and {Young}, Sam},
        title = {Primordial black hole formation during the QCD phase transition: Threshold, mass distribution, and abundance},
      journal = {PRD},
         year = {2024},
        month = {apr},
       volume = {109},
       number = {8},
        pages = {083506},
}

@ARTICLE{Mould2025,
       author = {{Mould}, Jeremy},
        title = {Dark Matter Genesis},
      journal = {ApJ},
         year = {2025},
        month = {may},
       volume = {984},
       number = {1},
          eid = {59},
        pages = {59},
archivePrefix = {arXiv},
       eprint = {2504.11595},
 primaryClass = {astro-ph.CO},
}

@ARTICLE{Farooq2025,
       author = {{Farooq}, Owais and {Zahoor}, Romana and {Francis}, Balungi},
        title = {Press-Schechter Formalism and The PBH Mass Distributions},
      journal = {arXiv e-prints},
         year = {2025},
        month = {jan},
          eid = {arXiv:2502.05194},
        pages = {arXiv:2502.05194},
          doi = {10.48550/arXiv.2502.05194},
archivePrefix = {arXiv},
       eprint = {2502.05194},
 primaryClass = {astro-ph.CO},
}

@article{kp23,
author={Korwar, M. and Profumo, S.},
title={Updated constraints on primordial black hole evaporation},
Year={2023},
Journal={JCAP}, 
Volume={05},
Pages={054}
}

@article{page76,
author={Page, D.},
year={1976},
journal={PhRvD},
volume={14},
pages={3260},
title={Particle emission rates from a black hole. II. Massless particles from a rotating hole}
}

@article{jane90,
author={MacGibbon, J. and Webber, B.},
year={1990},
journal={PhRvD.},
volume={41},
pages={3052},
title="{Quark- and gluon-jet emission from primordial black holes: The instantaneous spectra}"
}

@article{dh2020,
author={Liu, H. and Ridgway, G. and Slatyer, T.},
year={2020},
title={Code package for calculating modified cosmic ionization and thermal histories with dark matter and other exotic energy injections},
journal={PhRvD},
volume={101},
pages={3530}
}

@article{spf09,
author={Slatyer, T. and Padmanabhan, N. and Finkbeiner, D.},
title="{CMB constraints on WIMP annihilation: Energy absorption during the recombination epoch}",
year={2009},
journal={PhRvD},
volume={80},
pages={3526}
}

@article{yuan2026,
author={Guan-Wen Yuan and Marco Calz\'a and Davide Pedrotti},
title={Code package for calculating modified cosmic ionization and thermal histories with dark matter and other exotic energy injections},
journal={arxiv},
year={2025},
volume={2504.18270}
}

@article{calza,
author={Calz\'a, M.}, 
title={GrayHawk: A public code for calculating the Gray Body Factors of massless fields around spherically symmetric Black Holes},
journal={Phys. Dark Univ.},
volume={480},
pages={101900},
year={2025},
}

@article{p77,
author={Page, D.},
title={Particle emission rates from a black hole. III. Charged leptons from a nonrotating hole},
year={1976},
journal={Physical Review D},
Volume={16}, 
pages={2402}
}

@article{czl,
author={Chen, S. and Zhang, H.-H. and Long, G.}, 
year={2022},
title={Revisiting the constraints on primordial black hole abundance with the isotropic gamma-ray background},
journal={PhysRevD},
volume={105},
pages={063008}
}

@article{sl2016,
author={Slatyer, T.},
title={Indirect dark matter signatures in the cosmic dark ages. II. Ionization, heating, and photon production from arbitrary energy injections},
year={2016},
journal={PhRvD},
volume={93},
pages={3521},
}

@article{laha,
author={Laha, R.},
year={2019},
journal={PhRvL},
title={Primordial Black Holes as a Dark Matter Candidate Are Severely Constrained by the Galactic Center 511 keV  Line},
volume={123},
pages={1101}
}

@article{bc2019,
author={Boudaud, M. and Cirelli, M.},
title={Voyager 1, Further Constrain Primordial Black Holes as Dark Matter},
year={2019},
journal={PhRvL},
volume={122},
pages={1104}
}

@ARTICLE{Mosbech2022,
       author = {{Mosbech}, Markus and {Picker}, Zachary},
        title = "{Effects of Hawking evaporation on PBH distributions}",
      journal = {SciPost Physics},
         year = {2022},
        month = {oct},
       volume = {13},
       number = {4},
          eid = {100},
        pages = {100},
          doi = {10.21468/SciPostPhys.13.4.100},
archivePrefix = {arXiv},
       eprint = {2203.05743},
 primaryClass = {astro-ph.HE},
}

@ARTICLE{Page1976,
       author = {{Page}, D.~N. and {Hawking}, S.~W.},
        title = "{Gamma rays from primordial black holes.}",
      journal = {ApJ},
         year = {1976},
        month = {may},
       volume = {206},
        pages = {1-7},
}

@ARTICLE{Dixon2025,
       author = {{Dixon}, M. and {Mould}, J. and {Lidman}, C. and {Taylor}, E.~N. and {Flynn}, C. and {Duffy}, A.~R. and {Galbany}, L. and {Scolnic}, D. and {Davis}, T.~M. and {M{\"o}ller}, A. and {Kelsey}, L. and {Lee}, J. and {Wiseman}, P. and {Vincenzi}, M. and {Shah}, P. and {Aguena}, M. and {Allam}, S.~S. and {Alves}, O. and {Bacon}, D. and {Bocquet}, S. and {Brooks}, D. and {Burke}, D.~L. and {Carnero Rosell}, A. and {Carollo}, D. and {Carretero}, J. and {Conselice}, C. and {da Costa}, L.~N. and {Pereira}, M.~E.~S. and {Diehl}, H.~T. and {Doel}, P. and {Everett}, S. and {Ferrero}, I. and {Flaugher}, B. and {Frieman}, J. and {Garc{\'\i}a-Bellido}, J. and {Gatti}, M. and {Gaztanaga}, E. and {Giannini}, G. and {Gruen}, D. and {Gruendl}, R.~A. and {Gutierrez}, G. and {Herner}, K. and {Hinton}, S.~R. and {Hollowood}, D.~L. and {Honscheid}, K. and {James}, D.~J. and {Kuehn}, K. and {Lima}, M. and {Marshall}, J.~L. and {Mena-Fern{\'a}ndez}, J. and {Menanteau}, F. and {Miquel}, R. and {Myles}, J. and {Nichol}, R.~C. and {Ogando}, R.~L.~C. and {Palmese}, A. and {Pieres}, A. and {Plazas Malag{\'o}n}, A.~A. and {Samuroff}, S. and {Sanchez}, E. and {Sanchez Cid}, D. and {Sevilla-Noarbe}, I. and {Smith}, M. and {Sobreira}, F. and {Suchyta}, E. and {Swanson}, M.~E.~C. and {Tarle}, G. and {To}, C. and {Tucker}, B.~E. and {Tucker}, D.~L. and {Vikram}, V. and {Walker}, A.~R. and {Weaverdyck}, N.},
        title = {Calibrating the absolute magnitude of type Ia supernovae in nearby galaxies using [OII] and implications for H$_{0}$},
      journal = {mnras},
         year = {2025},
        month = {feb},
archivePrefix = {arXiv},
       eprint = {2408.01001},
 primaryClass = {astro-ph.CO},
}

@article{groupm,
	journal={arxiv},
	year={2025},
	title={Modified recombination and the Hubble tension},
		volume={2411.16678},
	author={ Mirpoorian, S.  and Jedamzik, K. and Pogosian, L.}
}

@ARTICLE{Planck2020,
author = {{Planck Collaboration} and {Aghanim}, N. and {Akrami}, Y. and {Ashdown}, M. and {Aumont}, J. and {Baccigalupi}, C. and {Ballardini}, M. and {Banday}, A.~J. and {Barreiro}, R.~B. and {Bartolo}, N. and {Basak}, S. and {Battye}, R. and {Benabed}, K. and {Bernard}, J. -P. and {Bersanelli}, M. and {Bielewicz}, P. and {Bock}, J.~J. and {Bond}, J.~R. and {Borrill}, J. and {Bouchet}, F.~R. and {Boulanger}, F. and {Bucher}, M. and {Burigana}, C. and {Butler}, R.~C. and {Calabrese}, E. and {Cardoso}, J. -F. and {Carron}, J. and {Challinor}, A. and {Chiang}, H.~C. and {Chluba}, J. and {Colombo}, L.~P.~L. and {Combet}, C. and {Contreras}, D. and {Crill}, B.~P. and {Cuttaia}, F. and {de Bernardis}, P. and {de Zotti}, G. and {Delabrouille}, J. and {Delouis}, J. -M. and {Di Valentino}, E. and {Diego}, J.~M. and {Dor{\'e}}, O. and {Douspis}, M. and {Ducout}, A. and {Dupac}, X. and {Dusini}, S. and {Efstathiou}, G. and {Elsner}, F. and {En{\ss}lin}, T.~A. and {Eriksen}, H.~K. and {Fantaye}, Y. and {Farhang}, M. and {Fergusson}, J. and {Fernandez-Cobos}, R. and {Finelli}, F. and {Forastieri}, F. and {Frailis}, M. and {Fraisse}, A.~A. and {Franceschi}, E. and {Frolov}, A. and {Galeotta}, S. and {Galli}, S. and {Ganga}, K. and {G{\'e}nova-Santos}, R.~T. and {Gerbino}, M. and {Ghosh}, T. and {Gonz{\'a}lez-Nuevo}, J. and {G{\'o}rski}, K.~M. and {Gratton}, S. and {Gruppuso}, A. and {Gudmundsson}, J.~E. and {Hamann}, J. and {Handley}, W. and {Hansen}, F.~K. and {Herranz}, D. and {Hildebrandt}, S.~R. and {Hivon}, E. and {Huang}, Z. and {Jaffe}, A.~H. and {Jones}, W.~C. and {Karakci}, A. and {Keih{\"a}nen}, E. and {Keskitalo}, R. and {Kiiveri}, K. and {Kim}, J. and {Kisner}, T.~S. and {Knox}, L. and {Krachmalnicoff}, N. and {Kunz}, M. and {Kurki-Suonio}, H. and {Lagache}, G. and {Lamarre}, J. -M. and {Lasenby}, A. and {Lattanzi}, M. and {Lawrence}, C.~R. and {Le Jeune}, M. and {Lemos}, P. and {Lesgourgues}, J. and {Levrier}, F. and {Lewis}, A. and {Liguori}, M. and {Lilje}, P.~B. and {Lilley}, M. and {Lindholm}, V. and {L{\'o}pez-Caniego}, M. and {Lubin}, P.~M. and {Ma}, Y. -Z. and {Mac{\'\i}as-P{\'e}rez}, J.~F. and {Maggio}, G. and {Maino}, D. and {Mandolesi}, N. and {Mangilli}, A. and {Marcos-Caballero}, A. and {Maris}, M. and {Martin}, P.~G. and {Martinelli}, M. and {Mart{\'\i}nez-Gonz{\'a}lez}, E. and {Matarrese}, S. and {Mauri}, N. and {McEwen}, J.~D. and {Meinhold}, P.~R. and {Melchiorri}, A. and {Mennella}, A. and {Migliaccio}, M. and {Millea}, M. and {Mitra}, S. and {Miville-Desch{\^e}nes}, M. -A. and {Molinari}, D. and {Montier}, L. and {Morgante}, G. and {Moss}, A. and {Natoli}, P. and {N{\o}rgaard-Nielsen}, H.~U. and {Pagano}, L. and {Paoletti}, D. and {Partridge}, B. and {Patanchon}, G. and {Peiris}, H.~V. and {Perrotta}, F. and {Pettorino}, V. and {Piacentini}, F. and {Polastri}, L. and {Polenta}, G. and {Puget}, J. -L. and {Rachen}, J.~P. and {Reinecke}, M. and {Remazeilles}, M. and {Renzi}, A. and {Rocha}, G. and {Rosset}, C. and {Roudier}, G. and {Rubi{\~n}o-Mart{\'\i}n}, J.~A. and {Ruiz-Granados}, B. and {Salvati}, L. and {Sandri}, M. and {Savelainen}, M. and {Scott}, D. and {Shellard}, E.~P.~S. and {Sirignano}, C. and {Sirri}, G. and {Spencer}, L.~D. and {Sunyaev}, R. and {Suur-Uski}, A. -S. and {Tauber}, J.~A. and {Tavagnacco}, D. and {Tenti}, M. and {Toffolatti}, L. and {Tomasi}, M. and {Trombetti}, T. and {Valenziano}, L. and {Valiviita}, J. and {Van Tent}, B. and {Vibert}, L. and {Vielva}, P. and {Villa}, F. and {Vittorio}, N. and {Wandelt}, B.~D. and {Wehus}, I.~K. and {White}, M. and {White}, S.~D.~M. and {Zacchei}, A. and {Zonca}, A.},
title = {Planck 2018 results. VI. Cosmological parameters},
journal = {aap},
year = {2020},
month = {sep},
volume = {641},
eid = {A6},
pages = {A6},
}

@Article{matplotlib2007, 
Author = {Hunter, J. D.}, 
Title = {Matplotlib: A 2D graphics environment}, 
Journal = {Computing In Science \& Engineering}, 
Volume = {9},
year={2007},
Number = {3}, 
Pages = {90--95}, 
}

@ARTICLE{Astropy2018, 
author = {Astropy Collaboration},
title = {Astropy: A community Python package for astronomy}, 
journal = {aap}, 
archivePrefix = "arXiv", 
eprint = {1307.6212}, 
primaryClass = "astro-ph.IM", 
year = 2013, 
month = oct, 
volume = 558, 
eid = {A33}, 
pages = {A33}, 
}

@ARTICLE{Astropy2013, 
author = {Astropy Collaboration}, 
title = {Astropy: A community Python package for astronomy}, 
journal = {aap}, 
archivePrefix = "arXiv", 
eprint = {1307.6212}, 
primaryClass = "astro-ph.IM", 
year = 2013, 
month = oct, 
volume = 558, 
eid = {A33}, 
pages = {A33}, 
}

@article{numpy2011, 
title={The NumPy array: a structure for efficient numerical computation}, author={van der Walt, Stefan and Colbert, S Chris and Varoquaux, Gael}, journal={Computing in Science \& Engineering}, volume={13}, number={2}, pages={22--30}, year={2011}, publisher={AIP Publishing} }

@Article{numpy2020,
 title         = {Array programming with {NumPy}},
 author        = {Charles R. Harris and K. Jarrod Millman and St{'{e}}fan J. van der Walt and Ralf Gommers and Pauli Virtanen and David Cournapeau and Eric Wieser and Julian Taylor and Sebastian Berg and Nathaniel J. Smith and Robert Kern and Matti Picus and Stephan Hoyer and Marten H. van Kerkwijk and Matthew Brett and Allan Haldane and Jaime Fern{'{a}}ndez del R{'{\i}}o and Mark Wiebe and Pearu Peterson and Pierre G{'{e}}rard-Marchant and Kevin Sheppard and Tyler Reddy and Warren Weckesser and Hameer Abbasi and Christoph Gohlke and Travis E. Oliphant},
 year          = {2020},
 month         = {sep},
 journal       = {Nature},
 volume        = {585},
 number        = {7825},
 pages         = {357--362},
 doi           = {10.1038/s41586-020-2649-2},
 publisher     = {Springer Science and Business Media {LLC}},
 url           = {https://doi.org/10.1038/s41586-020-2649-2}
}

@ARTICLE{CMasher2020,
author = {{van der Velden}, Ellert},
title = {CMasher: Scientific colormaps for making accessible, informative and 'cmashing' plots},
journal = {The Journal of Open Source Software},
year = {2020},
month = {Feb},
volume = {5},
number = {46},
eid = {2004},
pages = {2004},
}

@MISC{CMasher2024,
       author = {{van der Velden}, Ellert and {Robert}, Cl{\'e}ment and {Batten}, Adam and {Clauss}, Christian and {beskep} and {Haoyu}, ''Daniel'' and {Thyng}, Kristen},
        title = "{1313e/CMasher: v1.8.0}",
         year = {2024},
        month = {feb},
          eid = {10.5281/zenodo.10677366},
          doi = {10.5281/zenodo.10677366},
      version = {v1.8.0},
    publisher = {Zenodo},
}

@ARTICLE{Zeldovich1967,
       author = {{Zel'dovich}, Ya. B. and {Novikov}, I.~D.},
        title = {The Hypothesis of Cores Retarded during Expansion and the Hot Cosmological Model},
      journal = {sovast},
         year = {1967},
        month = {feb},
       volume = {10},
        pages = {602},
}

@ARTICLE{Hawking1971,
       author = {{Hawking}, Stephen},
        title = {Gravitationally collapsed objects of very low mass},
      journal = {mnras},
         year = {1971},
        month = {jan},
       volume = {152},
        pages = {75},
}

@article{ACT,
author= {Thibaut, L. and Adrien La Posta and Zachary Atkins and Hidde T. Jense and Irene Abril-Cabezas and Graeme E. Addison and Peter A.R. Ade and Simone Aiola and Tommy Alford and David Alonso },
title={The Atacama Cosmology Telescope: DR6 power spectra, likelihoods and ΛCDM parameters},
journal = {JCAP},
year = {2025},
volume= {11},
pages = {062},
}

@ARTICLE{Harada2013,
       author = {{Harada}, Tomohiro and {Yoo}, Chul-Moon and {Kohri}, Kazunori},
        title = {Threshold of primordial black hole formation},
      journal = {PRD},
         year = {2013},
        month = {oct},
       volume = {88},
       number = {8},
          eid = {084051},
        pages = {084051},
archivePrefix = {arXiv},
       eprint = {1309.4201},
 primaryClass = {astro-ph.CO},
}

@ARTICLE{Riess2022,
       author = {{Riess}, Adam G. and {Yuan}, Wenlong and {Macri}, Lucas M. and {Scolnic}, Dan and {Brout}, Dillon and {Casertano}, Stefano and {Jones}, David O. and {Murakami}, Yukei and {Anand}, Gagandeep S. and {Breuval}, Louise and {Brink}, Thomas G. and {Filippenko}, Alexei V. and {Hoffmann}, Samantha and {Jha}, Saurabh W. and {D'arcy Kenworthy}, W. and {Mackenty}, John and {Stahl}, Benjamin E. and {Zheng}, WeiKang},
        title = {A Comprehensive Measurement of the Local Value of the Hubble Constant with 1 km s$^{-1}$ Mpc$^{-1}$ Uncertainty from the Hubble Space Telescope and the SH0ES Team},
      journal = {ApJl},
         year = {2022},
        month = {jul},
       volume = {934},
       number = {1},
          eid = {L7},
        pages = {L7},
archivePrefix = {arXiv},
       eprint = {2112.04510},
 primaryClass = {astro-ph.CO},
}

@ARTICLE{Vagnozzi2023,
author = {{Vagnozzi}, Sunny},
title = {Seven Hints That Early-Time New Physics Alone Is Not Sufficient to Solve the Hubble Tension},
journal = {Universe},
year = {2023},
month = {aug},
volume = {9},
number = {9},
eid = {393},
pages = {393},
archivePrefix = {arXiv},
eprint = {2308.16628},
primaryClass = {astro-ph.CO},
}

@ARTICLE{DiValentino2021,
author = {{Di Valentino}, Eleonora and {Mena}, Olga and {Pan}, Supriya and {Visinelli}, Luca and {Yang}, Weiqiang and {Melchiorri}, Alessandro and {Mota}, David F. and {Riess}, Adam G. and {Silk}, Joseph},
title = {In the realm of the Hubble tension-a review of solutions},
journal = {Classical and Quantum Gravity},
year = {2021},
month = {jul},
volume = {38},
number = {15},
eid = {153001},
pages = {153001},
archivePrefix = {arXiv},
eprint = {2103.01183},
 primaryClass = {astro-ph.CO},
}

@ARTICLE{Vincenzi2024,
       author = {{Vincenzi}, M. and {Brout}, D. and {Armstrong}, P. and {Popovic}, B. and {Taylor}, G. and {Acevedo}, M. and {Camilleri}, R. and {Chen}, R. and {Davis}, T.~M. and {Lee}, J. and {Lidman}, C. and {Hinton}, S.~R. and {Kelsey}, L. and {Kessler}, R. and {M{\"o}ller}, A. and {Qu}, H. and {Sako}, M. and {Sanchez}, B. and {Scolnic}, D. and {Smith}, M. and {Sullivan}, M. and {Wiseman}, P. and {Asorey}, J. and {Bassett}, B.~A. and {Carollo}, D. and {Carr}, A. and {Foley}, R.~J. and {Frohmaier}, C. and {Galbany}, L. and {Glazebrook}, K. and {Graur}, O. and {Kovacs}, E. and {Kuehn}, K. and {Malik}, U. and {Nichol}, R.~C. and {Rose}, B. and {Tucker}, B.~E. and {Toy}, M. and {Tucker}, D.~L. and {Yuan}, F. and {Abbott}, T.~M.~C. and {Aguena}, M. and {Alves}, O. and {Allam}, S.~S. and {Andrade-Oliveira}, F. and {Annis}, J. and {Bacon}, D. and {Bechtol}, K. and {Bernstein}, G.~M. and {Brooks}, D. and {Burke}, D.~L. and {Carnero Rosell}, A. and {Carretero}, J. and {Castander}, F.~J. and {Conselice}, C. and {da Costa}, L.~N. and {Pereira}, M.~E.~S. and {Desai}, S. and {Diehl}, H.~T. and {Doel}, P. and {Ferrero}, I. and {Flaugher}, B. and {Friedel}, D. and {Frieman}, J. and {Garc{\'\i}a-Bellido}, J. and {Gatti}, M. and {Giannini}, G. and {Gruen}, D. and {Gruendl}, R.~A. and {Hollowood}, D.~L. and {Honscheid}, K. and {Huterer}, D. and {James}, D.~J. and {Kuropatkin}, N. and {Lahav}, O. and {Lee}, S. and {Lin}, H. and {Marshall}, J.~L. and {Mena-Fern{\'a}ndez}, J. and {Menanteau}, F. and {Miquel}, R. and {Palmese}, A. and {Pieres}, A. and {Plazas Malag{\'o}n}, A.~A. and {Porredon}, A. and {Romer}, A.~K. and {Roodman}, A. and {Sanchez}, E. and {Sanchez Cid}, D. and {Schubnell}, M. and {Sevilla-Noarbe}, I. and {Suchyta}, E. and {Swanson}, M.~E.~C. and {Tarle}, G. and {To}, C. and {Walker}, A.~R. and {Weaverdyck}, N. and {Yamamoto}, M. and {DES Collaboration}},
        title = {The Dark Energy Survey Supernova Program: Cosmological Analysis and Systematic Uncertainties},
      journal = {ApJ},
         year = {2024},
        month = {nov},
       volume = {975},
       number = {1},
          eid = {86},
        pages = {86},
          doi = {10.3847/1538-4357/ad5e6c},
archivePrefix = {arXiv},
       eprint = {2401.02945},
 primaryClass = {astro-ph.CO},
}

@ARTICLE{Perivolaropoulos2022,
       author = {{Perivolaropoulos}, L. and {Skara}, F.},
        title = {Challenges for {\ensuremath{\Lambda}}CDM: An update},
      journal = {nar},
         year = {2022},
        month = {dec},
       volume = {95},
          eid = {101659},
        pages = {101659},
archivePrefix = {arXiv},
       eprint = {2105.05208},
 primaryClass = {astro-ph.CO},
}

@ARTICLE{Bulla2022,
       author = {{Bulla}, Mattia and {Coughlin}, Michael W. and {Dhawan}, Suhail and {Dietrich}, Tim},
        title = {Multi-Messenger Constraints on the Hubble Constant Through Combination of Gravitational Waves, Gamma-Ray Bursts and Kilonovae from Neutron Star Mergers},
      journal = {Universe},
         year = {2022},
        month = {may},
       volume = {8},
       number = {5},
          eid = {289},
        pages = {289},
archivePrefix = {arXiv},
       eprint = {2205.09145},
 primaryClass = {astro-ph.HE},
}

@ARTICLE{Riess2024,
       author = {{Riess}, Adam G. and {Scolnic}, Dan and {Anand}, Gagandeep S. and {Breuval}, Louise and {Casertano}, Stefano and {Macri}, Lucas M. and {Li}, Siyang and {Yuan}, Wenlong and {Huang}, Caroline D. and {Jha}, Saurabh and {Murakami}, Yukei S. and {Beaton}, Rachael and {Brout}, Dillon and {Wu}, Tianrui and {Addison}, Graeme E. and {Bennett}, Charles and {Anderson}, Richard I. and {Filippenko}, Alexei V. and {Carr}, Anthony},
        title = {JWST Validates HST Distance Measurements: Selection of Supernova Subsample Explains Differences in JWST Estimates of Local H0},
      journal = {arXiv e-prints},
         year = {2024},
        month = {aug},
          eid = {arXiv:2408.11770},
        pages = {arXiv:2408.11770},
          doi = {10.48550/arXiv.2408.11770},
archivePrefix = {arXiv},
       eprint = {2408.11770},
 primaryClass = {astro-ph.CO},
}

@ARTICLE{Pascale2024,
       author = {{Pascale}, Massimo and {Frye}, Brenda L. and {Pierel}, Justin D.~R. and {Chen}, Wenlei and {Kelly}, Patrick L. and {Cohen}, Seth H. and {Windhorst}, Rogier A. and {Riess}, Adam G. and {Kamieneski}, Patrick S. and {Diego}, Jose M. and {Meena}, Ashish K. and {Cha}, Sangjun and {Oguri}, Masamune and {Zitrin}, Adi and {Jee}, M. James and {Foo}, Nicholas and {Leimbach}, Reagen and {Koekemoer}, Anton M. and {Conselice}, C.~J. and {Dai}, Liang and {Goobar}, Ariel and {Siebert}, Matthew R. and {Strolger}, Lou and {Willner}, S.~P.},
        title = {SN H0pe: The First Measurement of $H_0$ from a Multiply-Imaged Type Ia Supernova, Discovered by JWST},
      journal = {arXiv e-prints},
         year = {2024},
        month = {mar},
          eid = {arXiv:2403.18902},
        pages = {arXiv:2403.18902},
          doi = {10.48550/arXiv.2403.18902},
archivePrefix = {arXiv},
       eprint = {2403.18902},
 primaryClass = {astro-ph.CO},
}

@ARTICLE{Scolnic2023,
       author = {{Scolnic}, D. and {Riess}, A.~G. and {Wu}, J. and {Li}, S. and {Anand}, G.~S. and {Beaton}, R. and {Casertano}, S. and {Anderson}, R.~I. and {Dhawan}, S. and {Ke}, X.},
        title = {CATS: The Hubble Constant from Standardized TRGB and Type Ia Supernova Measurements},
      journal = {ApJl},
         year = {2023},
        month = {sep},
       volume = {954},
       number = {1},
          eid = {L31},
        pages = {L31},
archivePrefix = {arXiv},
       eprint = {2304.06693},
 primaryClass = {astro-ph.CO},
}

@ARTICLE{Kourkchi2020,
       author = {{Kourkchi}, Ehsan and {Tully}, R. Brent and {Anand}, Gagandeep S. and {Courtois}, H{\'e}l{\`e}ne M. and {Dupuy}, Alexandra and {Neill}, James D. and {Rizzi}, Luca and {Seibert}, Mark},
        title = "{Cosmicflows-4: The Calibration of Optical and Infrared Tully-Fisher Relations}",
      journal = {ApJ},
         year = {2020},
        month = {jun},
       volume = {896},
       number = {1},
          eid = {3},
        pages = {3},
archivePrefix = {arXiv},
       eprint = {2004.14499},
 primaryClass = {astro-ph.GA},
}

@ARTICLE{Schombert2020,
       author = {{Schombert}, James and {McGaugh}, Stacy and {Lelli}, Federico},
        title = "{Using the Baryonic Tully-Fisher Relation to Measure H$_{o}$}",
      journal = {AJ},
         year = {2020},
        month = {aug},
       volume = {160},
       number = {2},
          eid = {71},
        pages = {71},
          doi = {10.3847/1538-3881/ab9d88},
archivePrefix = {arXiv},
       eprint = {2006.08615},
 primaryClass = {astro-ph.CO},
}

@article{Pratt1960,
    author = "Pratt, R. H.",
    title = "{Atomic Photoelectric Effect at High Energies}",
    doi = "10.1103/PhysRev.117.1017",
    journal = "Phys. Rev.",
    volume = "117",
    pages = "1017--1028",
    year = "1960"
}

@article{Scofield1973,
author = {Scofield, J H},
title = {Theoretical photoionization cross sections from 1 to 1500 keV.},
doi = {10.2172/4545040},
url = {https://www.osti.gov/biblio/4545040}, journal = {},
year = {1973},
month = {1}
}

@article{Boccia2025,
author={Boccia, A. and Iocco, F. and Visinelli, L.},
year={2025},
title={Constraining the primordial black hole abundance through big-bang nucleosynthesis},
journal={Phys Rev D}, 
Volume={ 111}, 
pages={id.063508}
}

@ARTICLE{Slatyer2009,
       author = {{Slatyer}, Tracy R. and {Padmanabhan}, Nikhil and {Finkbeiner}, Douglas P.},
        title = "{CMB constraints on WIMP annihilation: Energy absorption during the recombination epoch}",
      journal = {PRD},
         year = {2009},
        month = {aug},
       volume = {80},
       number = {4},
          eid = {043526},
        pages = {043526},
archivePrefix = {arXiv},
       eprint = {0906.1197},
 primaryClass = {astro-ph.CO},
}

@ARTICLE{Seager1999,
       author = {{Seager}, S. and {Sasselov}, D.~D. and {Scott}, D.},
        title = "{A New Calculation of the Recombination Epoch}",
      journal = {ApJ},
         year = {1999},
        month = {sep},
       volume = {523},
       number = {1},
        pages = {L1-L5},
          doi = {10.1086/312250},
archivePrefix = {arXiv},
       eprint = {astro-ph/9909275},
 primaryClass = {astro-ph},
}

@ARTICLE{Seager2000,
       author = {{Seager}, Sara and {Sasselov}, Dimitar D. and {Scott}, Douglas},
        title = "{How Exactly Did the Universe Become Neutral?}",
      journal = {ApJS},
         year = {2000},
        month = {jun},
       volume = {128},
       number = {2},
        pages = {407-430},
          doi = {10.1086/313388},
archivePrefix = {arXiv},
       eprint = {astro-ph/9912182},
 primaryClass = {astro-ph},
}

@ARTICLE{Scott2009,
       author = {{Scott}, Douglas and {Moss}, Adam},
        title = "{Matter temperature during cosmological recombination}",
      journal = {mnras},
         year = {2009},
        month = {jul},
       volume = {397},
       number = {1},
        pages = {445-446},
archivePrefix = {arXiv},
       eprint = {0902.3438},
 primaryClass = {astro-ph.CO},
}

@ARTICLE{Hubbell1980,
       author = {{Hubbell}, J.~H. and {Gimm}, H.~A. and {{\O}verb{\o}}, I.},
        title = "{Pair, Triplet, and Total Atomic Cross Sections (and Mass Attenuation Coefficients) for 1 MeV-100 GeV Photons in Elements Z=1 to 100}",
      journal = {Journal of Physical and Chemical Reference Data},
         year = {1980},
        month = {oct},
       volume = {9},
       number = {4},
        pages = {1023-1148},
          doi = {10.1063/1.555629},
}

@ARTICLE{deJaeger2020,
       author = {{de Jaeger}, T. and {Stahl}, B.~E. and {Zheng}, W. and {Filippenko}, A.~V. and {Riess}, A.~G. and {Galbany}, L.},
        title = "{A measurement of the Hubble constant from Type II supernovae}",
      journal = {mnras},
         year = {2020},
        month = {aug},
       volume = {496},
       number = {3},
        pages = {3402-3411},
archivePrefix = {arXiv},
       eprint = {2006.03412},
 primaryClass = {astro-ph.CO},
}

@ARTICLE{Hamidreza2024,
       author = {{Hamidreza Mirpoorian}, Seyed and {Jedamzik}, Karsten and {Pogosian}, Levon},
        title = "{Modified recombination and the Hubble tension}",
      journal = {arXiv e-prints},
         year = {2024},
        month = {nov},
          eid = {arXiv:2411.16678},
        pages = {arXiv:2411.16678},
          doi = {10.48550/arXiv.2411.16678},
archivePrefix = {arXiv},
       eprint = {2411.16678},
 primaryClass = {astro-ph.CO},
}

@ARTICLE{Jedamzik2021,
       author = {{Jedamzik}, Karsten and {Pogosian}, Levon and {Zhao}, Gong-Bo},
        title = "{Why reducing the cosmic sound horizon alone can not fully resolve the Hubble tension}",
      journal = {Communications Physics},
         year = {2021},
        month = {dec},
       volume = {4},
       number = {1},
          eid = {123},
        pages = {123},
          doi = {10.1038/s42005-021-00628-x},
archivePrefix = {arXiv},
       eprint = {2010.04158},
 primaryClass = {astro-ph.CO},
}

@ARTICLE{Carr2021,
       author = {{Carr}, Bernard and {Kohri}, Kazunori and {Sendouda}, Yuuiti and {Yokoyama}, Jun'ichi},
        title = "{Constraints on primordial black holes}",
      journal = {Reports on Progress in Physics},
         year = {2021},
        month = {nov},
       volume = {84},
       number = {11},
          eid = {116902},
        pages = {116902},
          doi = {10.1088/1361-6633/ac1e31},
archivePrefix = {arXiv},
       eprint = {2002.12778},
 primaryClass = {astro-ph.CO},
}

@ARTICLE{Carr1974,
       author = {{Carr}, B.~J. and {Hawking}, S.~W.},
        title = "{Black holes in the early Universe}",
      journal = {mnras},
         year = {1974},
        month = {aug},
       volume = {168},
        pages = {399-416},
}

@ARTICLE{Poulter2019,
       author = {{Poulter}, Harry and {Ali-Ha{\"\i}moud}, Yacine and {Hamann}, Jan and {White}, Martin and {Williams}, Anthony G.},
        title = "{CMB constraints on ultra-light primordial black holes with extended mass distributions}",
      journal = {arXiv e-prints},
         year = {2019},
        month = {jul},
          eid = {arXiv:1907.06485},
        pages = {arXiv:1907.06485},
          doi = {10.48550/arXiv.1907.06485},
archivePrefix = {arXiv},
       eprint = {1907.06485},
 primaryClass = {astro-ph.CO},
}

@ARTICLE{Carr2010,
       author = {{Carr}, B.~J. and {Kohri}, Kazunori and {Sendouda}, Yuuiti and {Yokoyama}, Jun'Ichi},
        title = "{New cosmological constraints on primordial black holes}",
      journal = {PRD},
         year = {2010},
        month = {may},
       volume = {81},
       number = {10},
          eid = {104019},
        pages = {104019},
archivePrefix = {arXiv},
       eprint = {0912.5297},
 primaryClass = {astro-ph.CO},
}

@ARTICLE{Alam2017,
       author = {{Alam}, Shadab and {Ata}, Metin and {Bailey}, Stephen and {Beutler}, Florian and {Bizyaev}, Dmitry and {Blazek}, Jonathan A. and {Bolton}, Adam S. and {Brownstein}, Joel R. and {Burden}, Angela and {Chuang}, Chia-Hsun and {Comparat}, Johan and {Cuesta}, Antonio J. and {Dawson}, Kyle S. and {Eisenstein}, Daniel J. and {Escoffier}, Stephanie and {Gil-Mar{\'\i}n}, H{\'e}ctor and {Grieb}, Jan Niklas and {Hand}, Nick and {Ho}, Shirley and {Kinemuchi}, Karen and {Kirkby}, David and {Kitaura}, Francisco and {Malanushenko}, Elena and {Malanushenko}, Viktor and {Maraston}, Claudia and {McBride}, Cameron K. and {Nichol}, Robert C. and {Olmstead}, Matthew D. and {Oravetz}, Daniel and {Padmanabhan}, Nikhil and {Palanque-Delabrouille}, Nathalie and {Pan}, Kaike and {Pellejero-Ibanez}, Marcos and {Percival}, Will J. and {Petitjean}, Patrick and {Prada}, Francisco and {Price-Whelan}, Adrian M. and {Reid}, Beth A. and {Rodr{\'\i}guez-Torres}, Sergio A. and {Roe}, Natalie A. and {Ross}, Ashley J. and {Ross}, Nicholas P. and {Rossi}, Graziano and {Rubi{\~n}o-Mart{\'\i}n}, Jose Alberto and {Saito}, Shun and {Salazar-Albornoz}, Salvador and {Samushia}, Lado and {S{\'a}nchez}, Ariel G. and {Satpathy}, Siddharth and {Schlegel}, David J. and {Schneider}, Donald P. and {Sc{\'o}ccola}, Claudia G. and {Seo}, Hee-Jong and {Sheldon}, Erin S. and {Simmons}, Audrey and {Slosar}, An{\v{z}}e and {Strauss}, Michael A. and {Swanson}, Molly E.~C. and {Thomas}, Daniel and {Tinker}, Jeremy L. and {Tojeiro}, Rita and {Maga{\~n}a}, Mariana Vargas and {Vazquez}, Jose Alberto and {Verde}, Licia and {Wake}, David A. and {Wang}, Yuting and {Weinberg}, David H. and {White}, Martin and {Wood-Vasey}, W. Michael and {Y{\`e}che}, Christophe and {Zehavi}, Idit and {Zhai}, Zhongxu and {Zhao}, Gong-Bo},
        title = "{The clustering of galaxies in the completed SDSS-III Baryon Oscillation Spectroscopic Survey: cosmological analysis of the DR12 galaxy sample}",
      journal = {mnras},
         year = {2017},
        month = {sep},
       volume = {470},
       number = {3},
        pages = {2617-2652},
archivePrefix = {arXiv},
       eprint = {1607.03155},
 primaryClass = {astro-ph.CO},
}

@ARTICLE{Asgari2021,
       author = {{Asgari}, Marika and {Lin}, Chieh-An and {Joachimi}, Benjamin and {Giblin}, Benjamin and {Heymans}, Catherine and {Hildebrandt}, Hendrik and {Kannawadi}, Arun and {St{\"o}lzner}, Benjamin and {Tr{\"o}ster}, Tilman and {van den Busch}, Jan Luca and {Wright}, Angus H. and {Bilicki}, Maciej and {Blake}, Chris and {de Jong}, Jelte and {Dvornik}, Andrej and {Erben}, Thomas and {Getman}, Fedor and {Hoekstra}, Henk and {K{\"o}hlinger}, Fabian and {Kuijken}, Konrad and {Miller}, Lance and {Radovich}, Mario and {Schneider}, Peter and {Shan}, HuanYuan and {Valentijn}, Edwin},
        title = {KiDS-1000 cosmology: Cosmic shear constraints and comparison between two point statistics},
      journal = {aap},
         year = {2021},
        month = {jan},
       volume = {645},
          eid = {A104},
        pages = {A104},
archivePrefix = {arXiv},
       eprint = {2007.15633},
 primaryClass = {astro-ph.CO},
}

@ARTICLE{Abbott2018,
       author = {{Abbott}, T.~M.~C. and {Abdalla}, F.~B. and {Alarcon}, A. and {Aleksi{\'c}}, J. and {Allam}, S. and {Allen}, S. and {Amara}, A. and {Annis}, J. and {Asorey}, J. and {Avila}, S. and {Bacon}, D. and {Balbinot}, E. and {Banerji}, M. and {Banik}, N. and {Barkhouse}, W. and {Baumer}, M. and {Baxter}, E. and {Bechtol}, K. and {Becker}, M.~R. and {Benoit-L{\'e}vy}, A. and {Benson}, B.~A. and {Bernstein}, G.~M. and {Bertin}, E. and {Blazek}, J. and {Bridle}, S.~L. and {Brooks}, D. and {Brout}, D. and {Buckley-Geer}, E. and {Burke}, D.~L. and {Busha}, M.~T. and {Campos}, A. and {Capozzi}, D. and {Carnero Rosell}, A. and {Carrasco Kind}, M. and {Carretero}, J. and {Castander}, F.~J. and {Cawthon}, R. and {Chang}, C. and {Chen}, N. and {Childress}, M. and {Choi}, A. and {Conselice}, C. and {Crittenden}, R. and {Crocce}, M. and {Cunha}, C.~E. and {D'Andrea}, C.~B. and {da Costa}, L.~N. and {Das}, R. and {Davis}, T.~M. and {Davis}, C. and {De Vicente}, J. and {DePoy}, D.~L. and {DeRose}, J. and {Desai}, S. and {Diehl}, H.~T. and {Dietrich}, J.~P. and {Dodelson}, S. and {Doel}, P. and {Drlica-Wagner}, A. and {Eifler}, T.~F. and {Elliott}, A.~E. and {Elsner}, F. and {Elvin-Poole}, J. and {Estrada}, J. and {Evrard}, A.~E. and {Fang}, Y. and {Fernandez}, E. and {Fert{\'e}}, A. and {Finley}, D.~A. and {Flaugher}, B. and {Fosalba}, P. and {Friedrich}, O. and {Frieman}, J. and {Garc{\'\i}a-Bellido}, J. and {Garcia-Fernandez}, M. and {Gatti}, M. and {Gaztanaga}, E. and {Gerdes}, D.~W. and {Giannantonio}, T. and {Gill}, M.~S.~S. and {Glazebrook}, K. and {Goldstein}, D.~A. and {Gruen}, D. and {Gruendl}, R.~A. and {Gschwend}, J. and {Gutierrez}, G. and {Hamilton}, S. and {Hartley}, W.~G. and {Hinton}, S.~R. and {Honscheid}, K. and {Hoyle}, B. and {Huterer}, D. and {Jain}, B. and {James}, D.~J. and {Jarvis}, M. and {Jeltema}, T. and {Johnson}, M.~D. and {Johnson}, M.~W.~G. and {Kacprzak}, T. and {Kent}, S. and {Kim}, A.~G. and {King}, A. and {Kirk}, D. and {Kokron}, N. and {Kovacs}, A. and {Krause}, E. and {Krawiec}, C. and {Kremin}, A. and {Kuehn}, K. and {Kuhlmann}, S. and {Kuropatkin}, N. and {Lacasa}, F. and {Lahav}, O. and {Li}, T.~S. and {Liddle}, A.~R. and {Lidman}, C. and {Lima}, M. and {Lin}, H. and {MacCrann}, N. and {Maia}, M.~A.~G. and {Makler}, M. and {Manera}, M. and {March}, M. and {Marshall}, J.~L. and {Martini}, P. and {McMahon}, R.~G. and {Melchior}, P. and {Menanteau}, F. and {Miquel}, R. and {Miranda}, V. and {Mudd}, D. and {Muir}, J. and {M{\"o}ller}, A. and {Neilsen}, E. and {Nichol}, R.~C. and {Nord}, B. and {Nugent}, P. and {Ogando}, R.~L.~C. and {Palmese}, A. and {Peacock}, J. and {Peiris}, H.~V. and {Peoples}, J. and {Percival}, W.~J. and {Petravick}, D. and {Plazas}, A.~A. and {Porredon}, A. and {Prat}, J. and {Pujol}, A. and {Rau}, M.~M. and {Refregier}, A. and {Ricker}, P.~M. and {Roe}, N. and {Rollins}, R.~P. and {Romer}, A.~K. and {Roodman}, A. and {Rosenfeld}, R. and {Ross}, A.~J. and {Rozo}, E. and {Rykoff}, E.~S. and {Sako}, M. and {Salvador}, A.~I. and {Samuroff}, S. and {S{\'a}nchez}, C. and {Sanchez}, E. and {Santiago}, B. and {Scarpine}, V. and {Schindler}, R. and {Scolnic}, D. and {Secco}, L.~F. and {Serrano}, S. and {Sevilla-Noarbe}, I. and {Sheldon}, E. and {Smith}, R.~C. and {Smith}, M. and {Smith}, J. and {Soares-Santos}, M. and {Sobreira}, F. and {Suchyta}, E. and {Tarle}, G. and {Thomas}, D. and {Troxel}, M.~A. and {Tucker}, D.~L. and {Tucker}, B.~E. and {Uddin}, S.~A. and {Varga}, T.~N. and {Vielzeuf}, P. and {Vikram}, V. and {Vivas}, A.~K. and {Walker}, A.~R. and {Wang}, M. and {Wechsler}, R.~H. and {Weller}, J. and {Wester}, W. and {Wolf}, R.~C. and {Yanny}, B. and {Yuan}, F. and {Zenteno}, A. and {Zhang}, B. and {Zhang}, Y. and {Zuntz}, J.},
        title = "{Dark Energy Survey year 1 results: Cosmological constraints from galaxy clustering and weak lensing}",
      journal = {PRD},
         year = {2018},
        month = {aug},
       volume = {98},
       number = {4},
          eid = {043526},
        pages = {043526},
archivePrefix = {arXiv},
       eprint = {1708.01530},
 primaryClass = {astro-ph.CO},
       adsurl = {https://ui.adsabs.harvard.edu/abs/2018PhRvD..98d3526A}
}

@ARTICLE{Zhang2007,
       author = {{Zhang}, Le and {Chen}, Xuelei and {Kamionkowski}, Marc and {Si}, Zong-Guo and {Zheng}, Zheng},
        title = "{Constraints on radiative dark-matter decay from the cosmic microwave background}",
      journal = {PRD},
         year = {2007},
        month = {sep},
       volume = {76},
       number = {6},
          eid = {061301},
        pages = {061301},
archivePrefix = {arXiv},
       eprint = {0704.2444},
 primaryClass = {astro-ph},
}
